\documentclass[final,5p,times,twocolumn]{elsarticle}

\usepackage{amssymb}
\usepackage{subfig}
\usepackage{multirow}
\usepackage{diagbox}
\usepackage{balance}
\usepackage{graphicx} 
\newtheorem{remark}{Remark}
\usepackage{amsmath} 
\makeatletter

\newcommand{\Rmnum}[1]{\expandafter\@slowromancap\romannumeral #1@}
\usepackage{booktabs}
\usepackage{makecell}
\usepackage{textcomp}

\usepackage{etoolbox}
\usepackage{soul}

\usepackage[monochrome]{color}

\AtBeginEnvironment{tabular}{\tiny}

\newcommand{\ra}[1]{\renewcommand{\arraystretch}{#1}}
\makeatletter
\def\tagform@#1{%
  \maketag@@@{\color{blue}(\ignorespaces#1\unskip\@@italiccorr)}%
}
\makeatother

\biboptions{sort&compress}

\journal{Applied Energy}

\begin{document}

\begin{frontmatter}



\title{Integrating Physics-Based Modeling with Machine Learning for Lithium-Ion Batteries}


\author[inst1]{Hao Tu}

\affiliation[inst1]{organization={Department of Mechanical Engineering, University of Kansas},
            city={Lawrence},
            postcode={KS 66045}, 
            country={USA}}

\author[inst2]{Scott Moura}

\affiliation[inst2]{organization={Department of Civil and Environmental Engineering, University of California},
            city={Berkeley},
            postcode={CA 94720}, 
            country={USA}}

\author[inst3]{Yebin Wang}

\affiliation[inst3]{organization={Mitsubishi Electric Research Laboratories},
            city={Cambridge},
            postcode={MA 02139}, 
            country={USA}}



\author[inst1]{Huazhen Fang\corref{cor1}}
\ead{fang@ku.edu}
\cortext[cor1]{Corresponding author}
\begin{abstract}

Mathematical modeling of lithium-ion batteries (LiBs) is a primary challenge in advanced battery management. This paper proposes two new frameworks to integrate   physics-based models with machine learning to achieve high-precision modeling for LiBs. The frameworks are characterized by informing the machine learning model of the state information of the physical model, enabling a deep integration between physics and machine learning. Based on the frameworks, a series of hybrid models are constructed, through combining an electrochemical model and an equivalent circuit model, respectively, with a feedforward neural network. The hybrid models are relatively parsimonious in structure and can provide considerable   voltage predictive accuracy under a broad range of C-rates, as shown by extensive simulations and experiments. The study further expands to  conduct aging-aware hybrid modeling, leading to the design of a hybrid model conscious of the state-of-health to make prediction. The experiments show that the model has high  voltage  predictive accuracy throughout a LiB's cycle life.

\end{abstract}




\end{frontmatter}


\section{Introduction}
\label{sec:intro}
Lithium-ion batteries (LiBs) represent a key energy storage technology for our industry and society. Today, they not only power billions of consumer electronics devices, but also enable electrified transportation, smart grid, and renewable energy adoption to drive the world forward into a decarbonized energy future. The surging use of LiBs has led to ever-growing demands for higher operating performance and safety. Optimal operation of LiBs involves state estimation, control, and diagnosis, which all rely on accurate and efficient dynamic models of LiBs. Mathematical modeling of LiBs hence has attracted intense research interest in the past decade \cite{Krewer:JES:2018, Chaturvedi:ICSM:2010}. In this paper, we propose to integrate physics-based modeling with data-driven machine learning to develop a new breed of models that harness their respective merits. The proposed models will be shown to offer high voltage  predictive accuracy, computational efficiency and applicability to a broad range of C-rates. 

\subsection{Literature Review}
The literature includes two main types of physics-based LiB models, namely, electrochemical models and equivalent circuit models (ECMs). Electrochemical models use electrochemical principles to comprehensively characterize the electrochemical reactions, lithium-ion diffusion and concentration changes in the electrode/electrolyte, as well as various associated processes during charging/discharging of LiBs. A well-known electrochemical model is the Doyle-Fuller-Newman (DFN) model, which is broadly considered reliable and precise enough for almost all LiB management scenarios~\cite{Doyle:JES:1993,Lee:JES:2021}. Its accuracy yet comes with enormous computational complexity. This hence has motivated an incessant search for streamlined electrochemical models to balance between accuracy and computational costs. The single particle model (SPM) is one of the most parsimonious, which represents each electrode as a spherical particle and delineates lithium-ion intercalation and diffusion in the particles~\cite{Santhanagopalan:JPS:2006}. With its simplified structure, it is computationally fast but accurate only for low to medium C-rates (below 1 C-rate). Based on the SPM, there is a wide range of improved versions for higher accuracy under different conditions. They usually supplement the SPM with characterizations of thermal behavior~\cite{Guo:JES:2011, Tanim:DSMC:2015}, electrolyte dynamics~\cite{Rahimian:JPS:2013, Moura:TCST:2016, Han:JPS:2015, Li:JES:2017}, degradation physics~\cite{Li:AE:2018}, and stress buildup~\cite{Li:JES:2017}. Another important line of research lies in applying model order reduction methods to the DFN, SPM or other electrochemical models, with the aim of accelerating numerical computation~\cite{XIA:IFAC:2017, 5400816, Rahn:ECM:2007, Limoge:TCST:2018, Kehs:ACC:2014, Fang:CEP:2014, Cai:JES:2009}.

Differently, ECMs leverage electrical circuits, usually based on resistors, capacitors, and voltage sources, to capture LiBs' current/voltage dynamics in a physically interpretable way. Compared to electrochemical models, ECMs have greatly more parsimonious structures and simpler governing equations, thus advantageous for computation and conducive to real-time control, prediction, and simulation. Some widely used ECMs include the Rint model, the Thevenin model, and the Dual Polarization model~\cite{He:Energies:2011,Mousavi:RSER:2014,Plett:Vol1:2015}. Recent literature has expanded the development of ECMs toward better prediction accuracy. Some studies seek to account for the effects of hysteresis and temperature on  a LiB's electrical dynamics~\cite{Hu:JPS:2012, Li:TIE:2018, Lee:TIE:2018, LIN:JPS:2014,Bahramipanah:TPE:2017}. Others design new ECMs to approximate certain electrochemical models~\cite{Ning:TCST:2020,Tian:IECON:2018, GENG:EA:2021,Li:Eact:2019,Li:JPS:2021}. While ECMs have found increasing popularity, their structural simplicity restricts their accuracy, making them useful only for low to medium C-rates.

For all the aforementioned models, their effectiveness and fidelity  will decrease as a LiB ages, since many parameters of a model can change drastically with the LiB's state-of-health (SoH). This hence has stimulated research on aging-aware modeling, where electrochemical models~\cite{Ramadass:JES:2004,TIPPMANN:JPS:2014,LEGRAND:JPS:2014,YANG:JPS:2017, VONLUDERS:JPS:2019,Keil:JES:2020} or   ECMs~\cite{Li:TIE:2018,Alan:ICIT:2010,Perez:TVT:2017,Chen:TEC:2006,Kim:TEC:2011} are coupled with different aging or degradation mechanisms intrinsic to LiBs.

Besides physical modeling, extracting models from data directly has become appealing, as ubiquitous onboard sensing has increased data availability for today's LiB systems. Machine learning (ML) tools, such as neural networks (NNs)~\cite{Cap:TEC:2011} and support vector machine~\cite{Wang:ECM:2006}, have been used to learn battery models from measurement data. These ML models are black-box approximations of LiBs' dynamics. Bypassing the use of physical principles, they can be constructed from data conveniently in practice and sufficiently accurate if trained on rich, informative enough data. Meanwhile, data can help grasp various uncertain factors that affect a LiB cell's dynamic behaviors.  However, unlike physics-based models, pure ML models generally lack generalizability and risk producing physically unreasonable or incorrect prediction in  out-of-sample scenarios. Also, training them often requires large amounts of high-quality data, which may not  always be  possible.

A close inspection indicates that physical modeling and ML modeling are constructively complementary to each other.  On the one hand, physics-based models can offer physical interpretations of LiBs' dynamic behaviors and extrapolate to any  charging/discharging scenarios meeting model assumptions. However, they either require much computation, as in the case of the DFN, or have inadequate accuracy when the model assumptions are not satisfied---for instance, the SPM and ECMs, usually designed for low to medium C-rates, will poorly predict LiBs' dynamics at high C-rates. Besides, some physical parameters of these models, like the diffusion coefficients in electrochemical models and resistances in ECMs, are subject to change due to different operating conditions, such as temperature and LiB's aging. This will eventually result in model mismatch if these parameters are not corrected in time. On the other hand, ML-based modeling is able to extract complicated input-to-output relationships underlying data, especially those evading precise characterization by physical principles or suffering uncertainty. As another benefit, ML models, once after being trained on datasets, can run fast with only fixed computational costs. Based on the above, there is an emerging interest in hybrid physics-ML modeling for LiBs to combine the respective merits of the two modeling approaches. The study in~\cite{Refai:DSCC:2011} couples a one-dimensional electrochemical model with different kinds of NNs. In~\cite{Park:ACC:2017}, recurrent NNs are used to learn the residuals between a LiB's terminal voltage and the SPM's output voltage. In~\cite{Feng:JPS:2020}, a simplified SPM and a lumped thermal model are combined with an NN in series to predict the terminal voltage. These hybrid models have a similar underlying structure---an NN takes the current and output voltage of a physical model as its input, and predicts the residual or actual terminal voltage as its output. However, from a physical perspective, the mappings represented by such NNs do not effectively hold at the level of LiBs' dynamics. The NNs, and consequently the hybrid models, are often not accurate enough in prediction even if they can achieve satisfactory training accuracy. Therefore, while the present studies indicate a promise of hybrid modeling for LiBs, this subject is still underexplored to live up to its potential.

\begin{figure}[t]
\vspace{2mm}
    \centering
    \includegraphics[width = 0.5\textwidth,trim={7.4cm .5cm 7.8cm 4.0cm},clip]{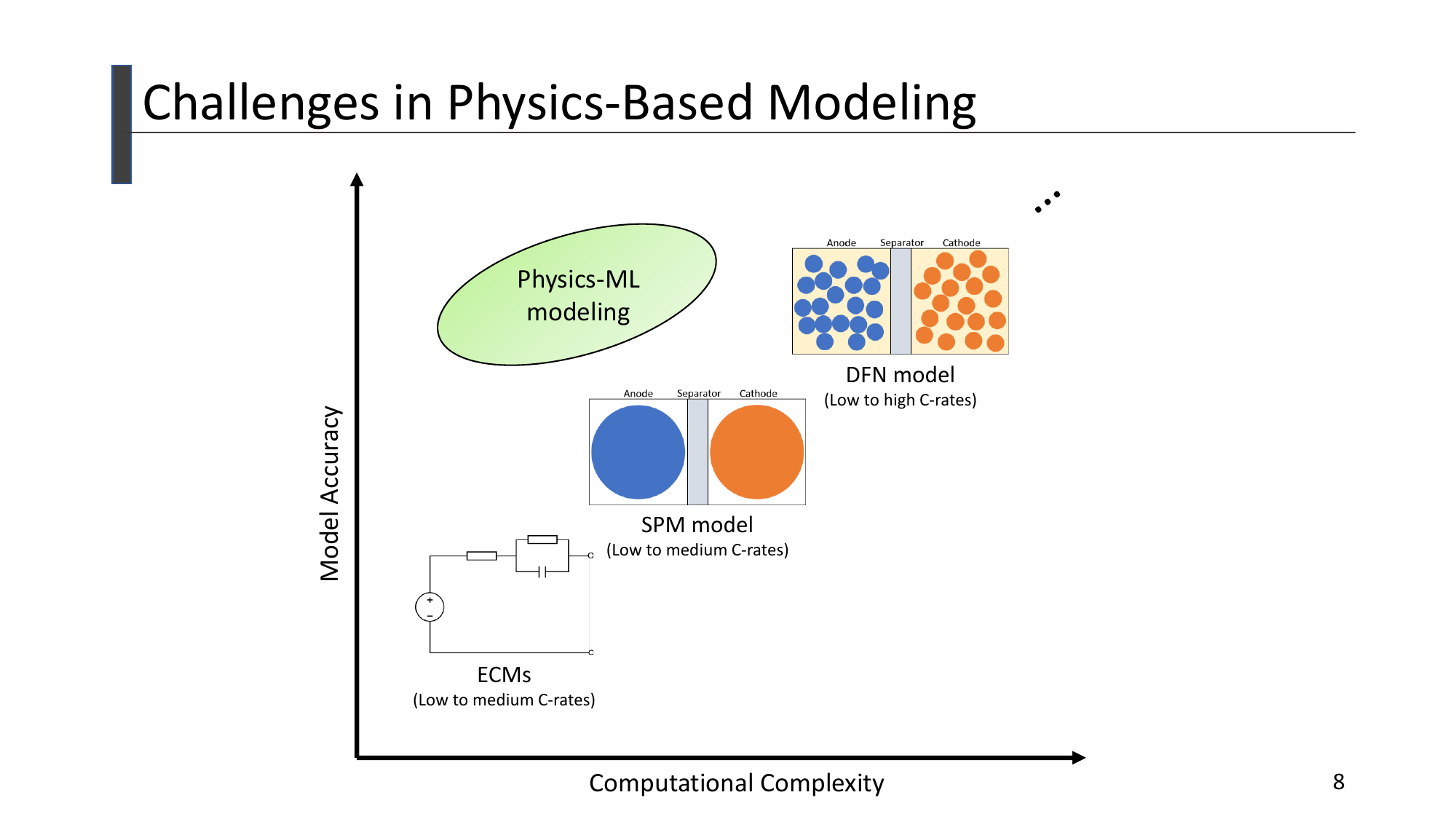}
    \caption{Comparison of physics-based models for LiB and their applicable current range.}
    \label{fig: Intro}
\end{figure}

\subsection{Contributions}

The goal of our study is to develop hybrid physics-ML models to enable highly accurate voltage prediction while preserving low computational complexity for LiBs, as visualized in Fig.~\ref{fig: Intro}. As pointed out in the literature survey, the existing hybrid models, e.g.,~\cite{Refai:DSCC:2011, Park:ACC:2017,Feng:JPS:2020}, use NNs to learn   relationships or mappings  that are not  physically meaningful, and thus see their predictive accuracy limited.  
 To overcome this   limitation, we propose a new perspective: {\em the NN must be informed of the internal state of the physical model to correctly learn what  the physical model misses in comparison to the actual physics of LiBs}. In other words,  the success of a hybrid model depends on whether the NN  represents a physically sound mapping; to this end, the NN must be made to take   the physical model's state as an input. 
The perspective leads us to develop the following specific contributions.
\begin{itemize}

\item We develop two   hybrid physics-ML LiB modeling frameworks, named as HYBRID-1 and HYBRID-2, respectively, which integrate physical LiB models with feedforward neural networks (FNNs). HYBRID-1 leverages an FNN to capture the residuals of a physical model, and HYBRID-2 uses an FNN to predict the terminal voltage based on a physical model. Different from the literature, both of them critically feed state information of the physical model to the FNN. In particular, we provide a mathematical reasoning to prove that  the designed frameworks are physically reasonable.

\item We apply  the above  frameworks to effectively integrate  electrochemical models and ECMs with FNNs. Our first effort combines the SPM with thermal dynamics (SPMT) with an FNN, and the second blends the nonlinear double capacitor (NDC) model, an ECM proposed recently in~\cite{Tian:IECON:2018, Ning:TCST:2020}, with an FNN. The developed models, first of their kind, are validated via extensive simulations or experiments, demonstrating high  voltage  predictive accuracy across broad C-rate ranges.

\item We further propose to incorporate aging awareness into hybrid modeling and develop an upgraded hybrid model that  utilizes a LiB cell's SoH information for voltage prediction. The model is shown capable of making accurate prediction throughout a cell's cycle life.

\end{itemize}

Compared to the existing hybrid models, the proposed frameworks and models can generalize and predict precisely beyond training datasets, thanks to the distinct attribute of making the FNN aware of the physical model's state. They may  find potential use in various LiB energy storage applications, especially those involving high C-rates and high power load conditions.  A further view  of their   applications is given in Section~\ref{Discussion}. 


\subsection{Organization}
This paper is organized as follows. Section 2 presents the two proposed hybrid modeling frameworks. Based on the frameworks, Sections 3-4 develop hybrid models based on integrating the SPMT and the NDC with FNNs, respectively, and validate them. Then, Section 5 constructs an aging-aware hybrid model based upon Sections 3-4 and verifies the results. Section 6 provides our remarks about the results. Finally, Section 7 concludes the study.


{\color{blue}A preliminary conference version of the work appeared in \cite{Tu:ACC:2021}, which deals with only the integration of electrochemical modeling with ML. This paper introduces significant extensions to improve the study in both depth and breadth. The extensions include the following: 1) the addition of a mathematical rationale to explain the correctness of the proposed hybrid modeling frameworks, 2) the development of new hybrid models by integrating an ECM with ML, and evaluation of them by experiments, and 3) the expansion of the proposed frameworks to aging-aware hybrid modeling along with experimental validation.}

\section{Hybrid Physics-ML Modeling for LiBs}
\label{sec:Hybrid Modeling}
In this section, we present two hybrid physics-ML modeling frameworks, referring to them as HYBRID-1 and HYBRID-2, respectively. They both are designed to blend physical modeling with an FNN, and their difference lies in the learning objective set for the FNN. We further provide an overview of FNNs for the sake of completeness.

\subsection{The Proposed Hybrid Modeling Frameworks}
\begin{figure}[t!]
\centering
\subfloat[][]{
    \includegraphics[width = .49\textwidth,trim={6.9cm 2cm 4.5cm 3.5cm},clip]{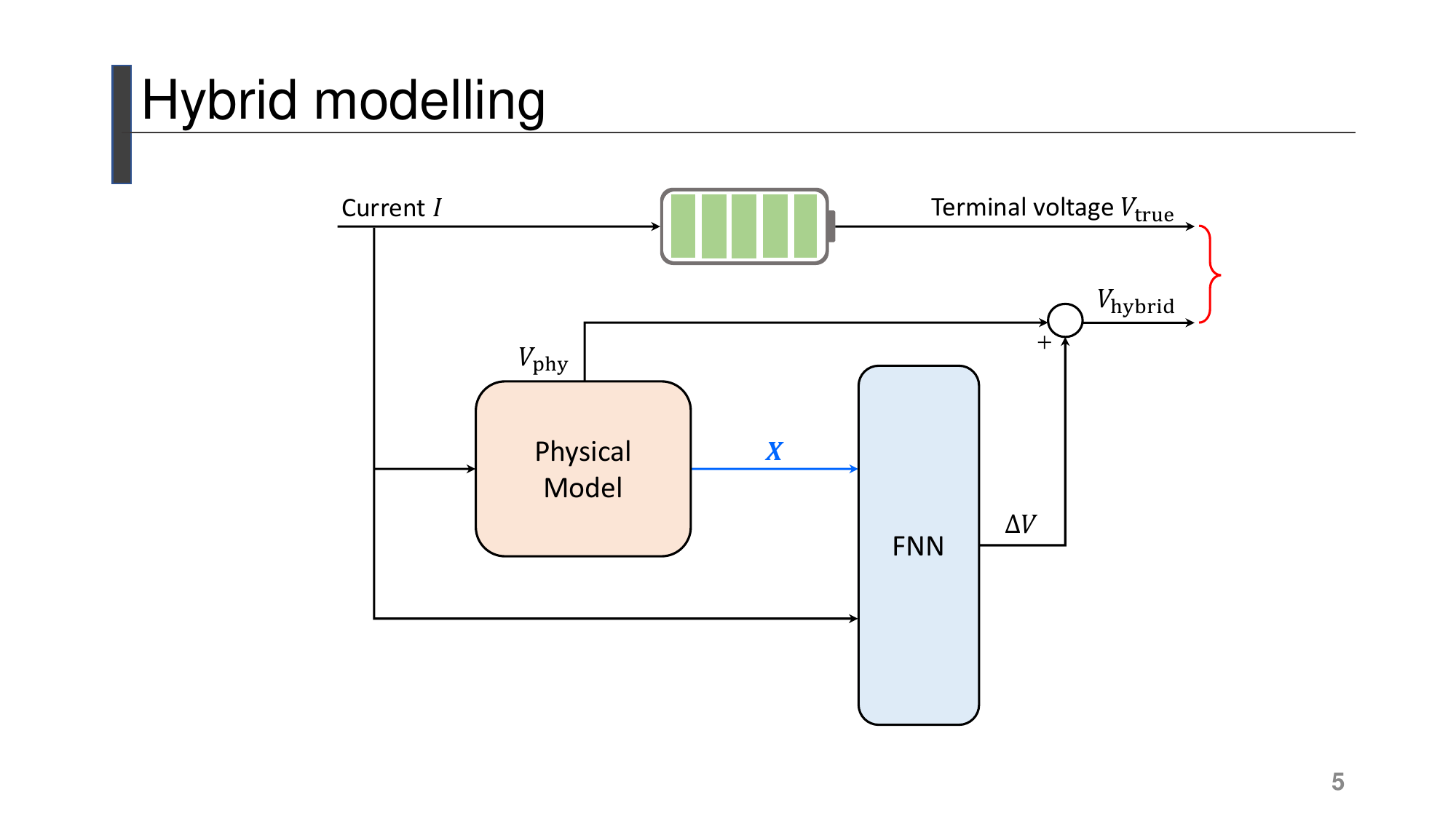}
    \label{fig: FRAME1}}
    
\subfloat[][]{
    \includegraphics[width = .49\textwidth,trim={5.5cm 4cm 6.1cm 3cm},clip]{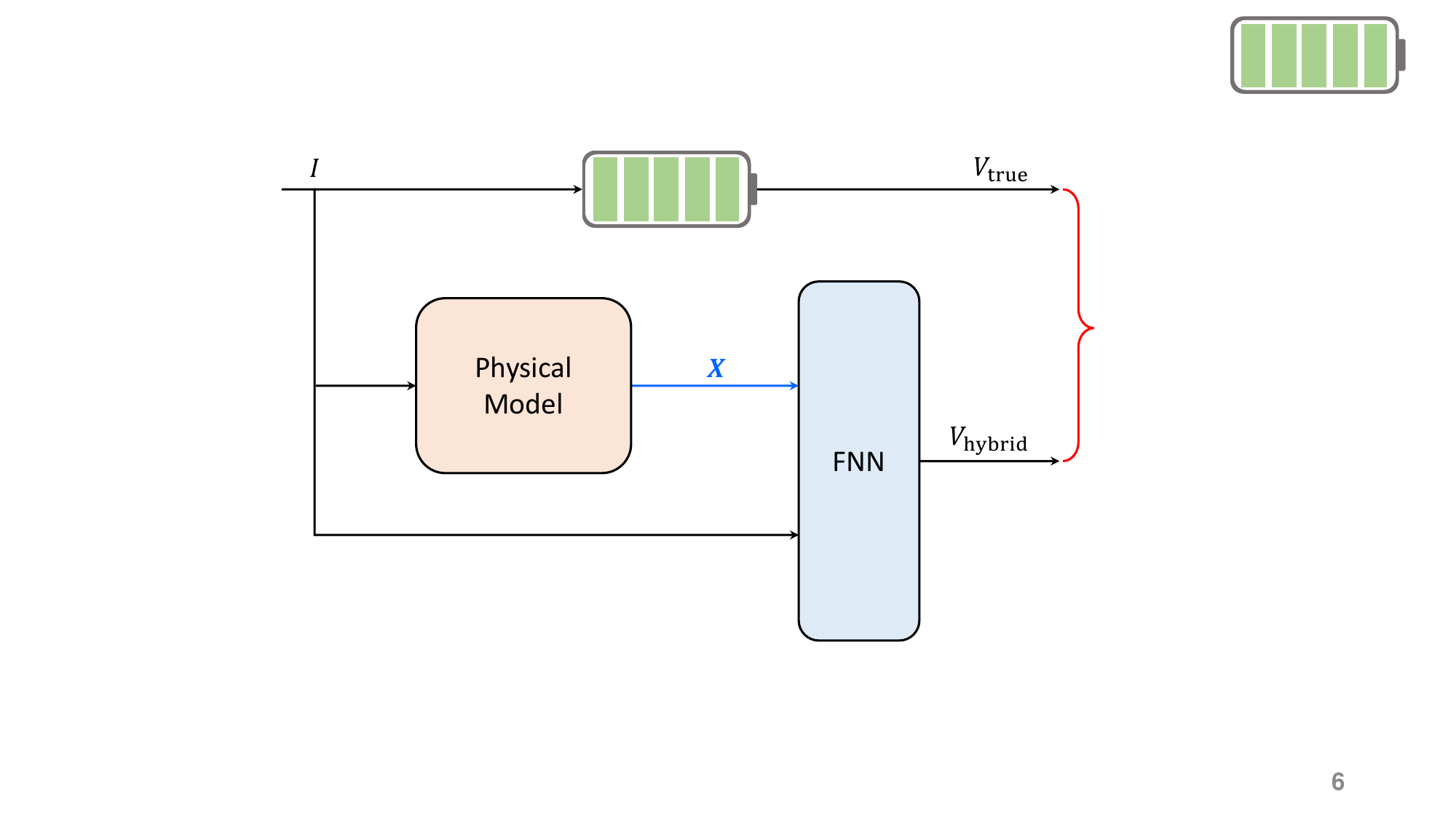}
    \label{fig: FRAME2}}
\caption{Block diagrams of (a) the HYBRID-1 framework and (b) the HYBRID-2 framework.}
\end{figure}

As shown in Fig.~\ref{fig: FRAME1}, HYBRID-1 is composed of a physical model in cascade with an FNN, with them operating simultaneously. The physical model approximately represents a LiB cell's electrochemical, electrical, or thermal behaviors. It is not perfectly accurate relative to the cell's true dynamics, due to inevitable model mismatch or uncertainty. The FNN   is used to learn biases of the physical model. Here, it is set to capture $\Delta V = V_\mathrm{true} - V_\mathrm{phy}$, which is the physical model's residual error with respect to the true terminal voltage. Leveraging the FNN's prediction $\Delta V$ to correct $V_\mathrm{phy}$, HYBRID-1 will output $V_\mathrm{hybrid} = V_\mathrm{phy} + \Delta V$ to emulate the cell's actual voltage. As an extension, we propose HYBRID-2  shown in Fig.~\ref{fig: FRAME2}, in which the FNN is made to learn the LiB's terminal voltage $V_\mathrm{true}$ directly, rather than the residual. By design, the FNN here is also informed of the state information of the physical model as in HYBRID-1.

It is critical to select the input variables of the FNN so that the FNN can learn correct relationships  consistent with the LiB cell's true dynamics. We propose that the FNN should take the physical model's state information $\boldsymbol{X}$ and the applied current $I$ as its input. The reasoning  is as follows. Without loss of generality, let us consider that the LiB's actual dynamics follows a high-dimensional nonlinear model of the form
\begin{align}\label{HD-nonlinear-sys}
\left\{
\begin{aligned}
\dot{\boldsymbol{\xi}} &= f({\boldsymbol{\xi},I}),\\
{V}_{\mathrm{true}} &= h({\boldsymbol{\xi}}, {I}),
\end{aligned}
\right.
\end{align}
where ${\boldsymbol{\xi}}\in \mathbb{R}^p$ is the  full-order state. The model may be derived from the ordinary differential equations or discretization of the partial differential equations governing the LiB~\cite{XIA:IFAC:2017,Kehs:ACC:2014,Cai:JES:2009}. The physical model can be viewed as a reduced-order representation of the LiB's full   actual dynamics, which approximates the original model in~\eqref{HD-nonlinear-sys} as
\begin{align}\label{LD-nonlinear-sys}
\left\{
\begin{aligned}
\dot{\boldsymbol{X}} &= f_r (\boldsymbol{X}, I),\\
{V}_{\mathrm{phy}}  &= h_r(\boldsymbol{X}, I),
\end{aligned}
\right.
\end{align}
where $\boldsymbol{X} \in \mathbb{R}^q$ with $q \ll p$ is the reduced-order state. From the perspective of model order reduction, one can view $\boldsymbol{X}$ as the result of projecting the full state $\boldsymbol{\xi}$ into  a low-dimensional space. The projection can be described as $
\boldsymbol{X} = \sigma (\boldsymbol{\xi})$, where $\sigma:  \mathbb{R}^p \rightarrow \mathbb{R}^q$.  Note that it is not possible to exactly reconstruct $\boldsymbol{\xi}$ using $\boldsymbol{X}$. However, 
since both $\boldsymbol{\xi}$ and $\boldsymbol{X}$ represent or reflect the state of the same LiB, it is reasonable to assume that there exists a nonlinear transformation to approximately project  $\boldsymbol{X}$ back to $\boldsymbol{\xi}$:
\begin{align}\label{Inverse-trans}
\boldsymbol{\xi} =  \psi ( \boldsymbol{X} , I) + \boldsymbol{\epsilon},
\end{align}
where $\boldsymbol{\epsilon}$ is the approximation error. Then, according to~\eqref{HD-nonlinear-sys}-\eqref{Inverse-trans}, the residual $\Delta V$ can be expressed as
\begin{align*}
\Delta V &= V_{\mathrm{true}} - V_{\mathrm{phy}} \\ &= h(\psi ( \boldsymbol{X} , I) + \boldsymbol{\epsilon}, I) - h_r(\boldsymbol{X}, I) \\ &\approx h \left( \psi ( \boldsymbol{X} , I), I \right) - h_r( \boldsymbol{X}, I),
\end{align*}
where the approximate equality follows from the zeroth-order Taylor expansion of $h(\psi ( \boldsymbol{X} , I) + \boldsymbol{\epsilon}, I)$ around $\boldsymbol{X}$ and $\boldsymbol{\epsilon} = 0$. This implies an approximate mapping   $( \boldsymbol{X}, {I} )  \rightarrow \Delta V$. We hence can use  an FNN to learn this mapping as in the HYBRID-1 framework, with $(\boldsymbol{X}, {I})$ as the input and ${\Delta V}$ as the output of the FNN.  Following similar lines, we can find
\begin{align*}
V_\mathrm{true} \approx   h \left( \psi ( \boldsymbol{X} , I), I \right).
\end{align*}
This relation justifies using  an FNN to learn the approximate mapping $( \boldsymbol{X} , I)  \rightarrow V_\mathrm{true}$, as is done in the HYBRID-2 framework.

\begin{remark}
The pivotal difference of the above hybrid modeling design from the literature, e.g., ~\cite{Refai:DSCC:2011,Park:ACC:2017, Feng:JPS:2020}, is that information about the physical model's state is fed as part of the input to the FNN. This, as is reasoned above, makes the FNN capable of learning  physically consistent relationships, and the resultant tighter physics-ML integration will lead to enhanced accuracy in prediction.
\end{remark}

\begin{remark}
HYBRID-1 and HYBRID-2 are modular and extensible frameworks that allow execution in versatile ways to construct different hybrid models. First, one can use either an electrochemical model or an ECM as the physical model component, depending on the specific objective of hybrid modeling. To demonstrate this, we will exploit the SPMT model and the NDC model, respectively, in Sections 3-4. Further, the frameworks can be readily extended to meet more needs. For instance, in Section 5, we will incorporate the SoH information into the formulation, enabling the FNN to make prediction with an awareness of a LiB's aging condition. This improvement will lead to hybrid models being able to predict voltage dynamics throughout the LiB's cycle life. Finally, the frameworks are open to using other ML models, e.g., Gaussian processes or support vector machines, even though this study focuses on the FNN. 
\end{remark}

\begin{remark}
Physics-informed ML for battery modeling has attracted growing attention recently. Among the few   studies,   NNs are used in~\cite{LiW:JPS:2021} to estimate the internal states of a physical model, e.g., concentrations and potentials in the electrodes and the electrolyte, and in~\cite{NASCIMENTO:JPS:2021} to capture the variability in the non-ideal voltage  term  of an electrochemical  model. While these  are meaningful ways to enhance battery modeling, this paper pursues   a  different goal of using physics-informed ML for highly accurate voltage prediction over broad C-rate ranges. This type of battery modeling is  useful and important for a variety of battery management tasks, and its potential applications is further discussed in Section~\ref{Discussion}.
\end{remark}

\subsection{The FNN Model}

FNNs are an important class of ML methods designed to approximate complex functions. Their network structure contains no cycle or feedback connections, making them the simplest type of NNs and  easy to train and implement. The theoretical performance of FNNs is guaranteed by the universal approximation theorem, which generally states that a continuous vector-valued function in the real space can be approximated with arbitrary accuracy by an FNN~\cite{HORNIK1989359}. An overview of FNNs is offered below, which is mainly based on~\cite{Goodfellow:DeepLearningBook, Malmstrom:IFAC:2020}.

\begin{figure}[t!]
    \centering
    \includegraphics[width = 0.5\textwidth,trim={5cm 2.2cm 5cm .9cm},clip]{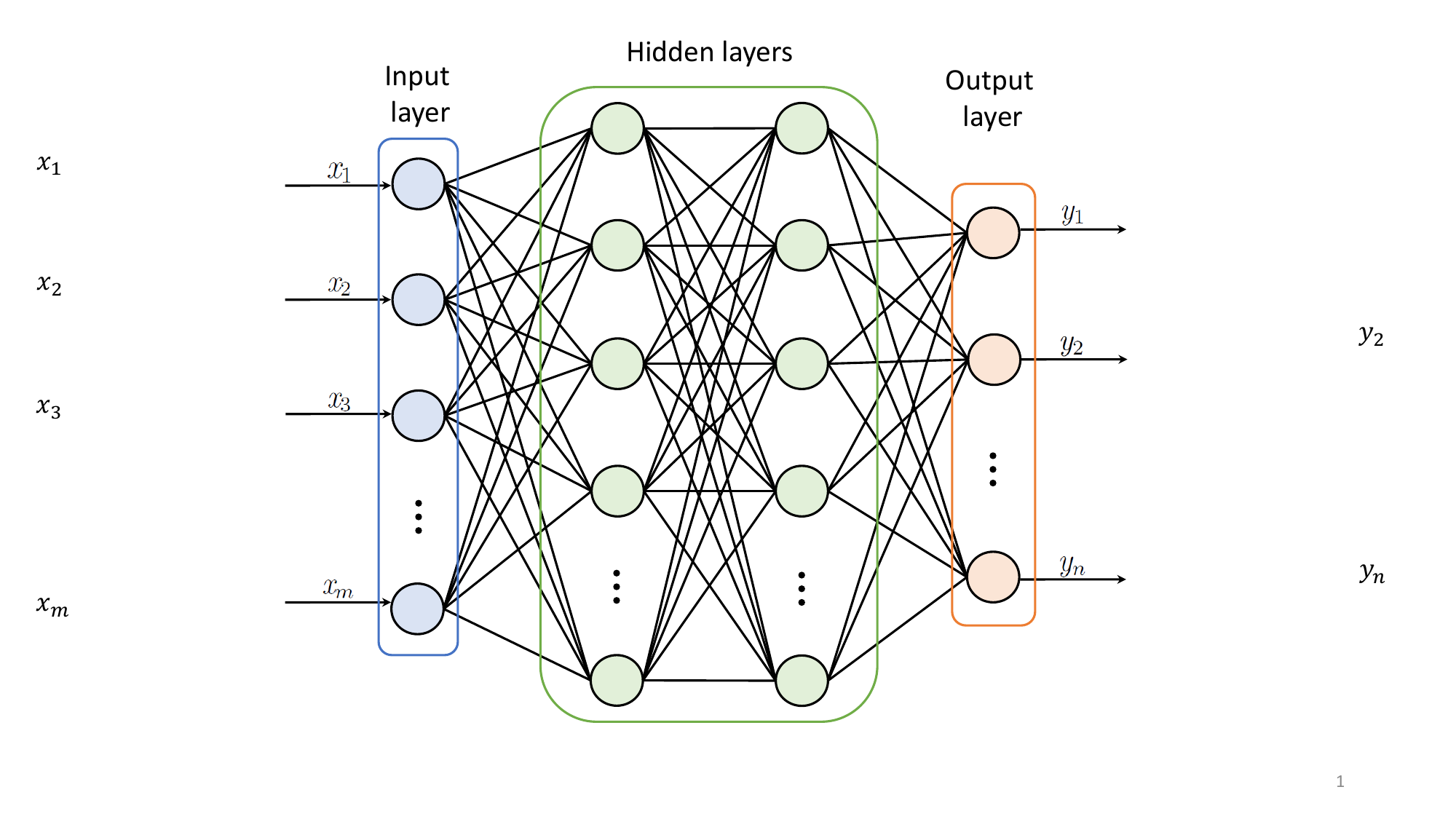}
    \caption{FNN architecture with two fully connected hidden layers.}
    \label{fig:NN}
\end{figure}

Consider an unknown function $g^\ast$, which is a mapping from an $m$-dimensional input $\boldsymbol  x$ to an $n$-dimensional output $\boldsymbol y$. An FNN aims to approximate it by defining  a parameterized mapping ${\boldsymbol y} = g\left(\boldsymbol x, \boldsymbol \theta \right)$ and learning the collection of parameters $\boldsymbol \theta$ from the data $\left\{ ({\boldsymbol x}_i, {\boldsymbol y}_i), i=1,2,\ldots,N \right\}$. As Fig.~\ref{fig:NN} shows, the structure of an FNN includes three parts interconnected in series, namely the input layer, hidden layers, and output layer. The input layer takes the input $\boldsymbol x$ and passes it to the first hidden layer. A hidden layer makes a nonlinear transformation of its input. For example, the first hidden layer  will transform $\boldsymbol x$ into $\phi(\boldsymbol W_1  \boldsymbol x + \boldsymbol b_1)$, where $\phi$ is a chosen nonlinear  mapping often called as activation function, $\boldsymbol W_1$ is the weight matrix, and $\boldsymbol b_1$ is a correction term. The following hidden layers then run similar nonlinear transformations sequentially. Finally, the output layer generates an output value to match $\boldsymbol y$. An $L$-layer FNN can be described in a general form: 
\begin{align*}
{\boldsymbol z}_1 &= {\boldsymbol x} ,\\
{\boldsymbol z}_{l } &= \phi \left(  {\boldsymbol W}_{l-1}  {\boldsymbol z}_{l-1} + {\boldsymbol b}_{l-1} \right), \ l =   2, 3, \ldots, L-1,\\
{\boldsymbol y} &= {\boldsymbol W}_{L-1}  {\boldsymbol z}_{L-1}  + {\boldsymbol b}_{L-1},
\end{align*}
where $ {\boldsymbol z}_{l-1}$ and  ${\boldsymbol z}_{l}$  are the input and output of the $l$-th layer, respectively. Note that the information flows only in the forward direction from $\boldsymbol x$ to $\boldsymbol y$  in the above network model, which is why the model is called as $feedforward$ NN. 

For the FNN, $\boldsymbol \theta$ is the collection of ${\boldsymbol W}_l$ and ${\boldsymbol b}_l$ for $l=1, 2, \ldots, L-1$. The training of the FNN is to identify $\boldsymbol \theta$ from  the measurement data $\left\{ ({\boldsymbol x}_i, {\boldsymbol y}_i)  \right\}$. A common approach is based on   maximum likelihood estimation, which minimizes the following cost function:
\begin{align*}
J(\boldsymbol \theta) = - \mathbb{E}_{{\boldsymbol x}, {\boldsymbol y} \sim \hat p_{\mathrm{data}} } \log p_{\mathrm{model}} \left( {\boldsymbol y} \mid {\boldsymbol x}, {\boldsymbol \theta}\right),
\end{align*}
where $\hat p_{\mathrm{data}}$ is the data-based empirical distribution of $\boldsymbol x$ and $\boldsymbol y$, and $p_{\mathrm{model}}$ is the probability distribution of $\boldsymbol y$ over the parameter space $\boldsymbol \theta$ based on the FNN model. Under some assumptions, $J(\boldsymbol \theta)$ can reduce to a mean squared error cost: 
\begin{align*}
J(\boldsymbol \theta)  =  \frac{1}{N} \sum_{i=1}^N \left\| {\boldsymbol y}_i - g \left(  {\boldsymbol x}_i, \boldsymbol \theta \right) \right\|^2.
\end{align*}
The minimization is usually achieved using stochastic gradient descent algorithms.  The computation of the gradient can be complicated, especially for multi-layer FNNs,  but it can still be done efficiently by the back-propagation algorithm or its generalizations.

\section{Hybrid Modeling {\em via} SPMT+FNN}
\label{sec:Electro Hybrid}

Based on the HYBRID-1 and HYBRID-2 frameworks, we integrate the SPMT model with an FNN to build two hybrid models, named as SPMTNet-1 and SPMTNet-2, respectively. The proposed models are validated {\em via} extensive simulations.

\subsection{The SPMTNet-1 and SPMTNet-2 Models}
\label{sec: The SPMTNets}

Developed in~\cite{Guo:JES:2011}, the SPMT model couples the SPM model with a bulk thermal model to predict the electrochemical and thermal behaviors simultaneously. The SPM simplifies each electrode of a LiB cell as a spherical particle and disregards the electrolyte dynamics. The transport of the lithium ions inside a  particle is governed by the Fick's diffusion law in spherical coordinates:
\begin{align}\label{SPM-Diffusion}
    \frac{\partial c_s^\pm}{\partial t}(r,t) = \frac{1}{r^2}\frac{\partial}{\partial r}\left[D_s^\pm r^2 \frac{\partial c_s^\pm}{\partial r}(r,t)\right] ,
\end{align}
where $c_s^\pm(r,t)$ is the solid-phase lithium-ion concentration of positive ($+$) or negative ($-$) electrode, and $D_s^\pm$ is the solid-phase diffusion coefficient. The boundary conditions of~\eqref{SPM-Diffusion} are given by 
\begin{align*}
    \frac{\partial c_s^\pm}{\partial r}(0,t) = 0 \ \ \ \mathrm{and} \ \ \ 
    \frac{\partial c_s^\pm}{\partial r}(R_s^\pm,t) = -\frac{1}{D_s^\pm} j_n^\pm ,
\end{align*}
where $R_s^\pm$ is the particle  radius and $j_n^\pm$ is the molar flux at the particle surface. Here,  
\begin{align*}
    j_n^\pm = \mp \frac{I(t)}{a_s^\pm F A L^\pm} ,
\end{align*}
where $a_s^{\pm}$ is the specific interfacial area, $F$ is the Faraday's constant, $A$ is an electrode's surface area, and $L^{\pm}$ is the electrode's thickness. Further, $j_n^\pm$ results from the electrochemical kinetics and depends on the overpotential of the electrodes $\eta^\pm$. The relation is  characterized by the Butler-Volmer equation:
\begin{align}\label{Eqn: B-V}
j_n^\pm = \frac{1}{F} i_0^\pm\left[ \exp\left({\frac{\alpha_{a}F}{RT}}\eta^{\pm}\right)-\exp\left({\frac{-\alpha_{c}F}{RT}\eta^{\pm}}\right) \right] .
\end{align}
Here, $\alpha_{a}$ and $\alpha_{c}$ are the anodic and cathodic charge transfer coefficients, respectively, and $i_0^\pm$ is the exchange current density given by
\begin{align*}
    i_0^\pm = k^\pm \left( c_e^{0} \right) ^{\alpha_{a}} \left( c_{ss}^\pm(t) \right) ^{\alpha_{c}} \left( c_{s,\max}^\pm - c_{ss}^\pm(t) \right) ^{\alpha_{a}},
\end{align*}
where $k^\pm$ is the kinetic reaction rate, $c_e^{0}$ is the constant electrolyte-phase lithium-ion concentration, $c_{ss}^\pm(t)=c_s^\pm\left(R_s^\pm,t\right)$ is the solid-phase lithium-ion concentration at the particle surface and $c_{s,\max}^\pm$ is the maximum solid-phase lithium-ion concentration. By assuming $\alpha_{a}=\alpha_{c}=0.5$, {\color{blue}(\ref{Eqn: B-V})} indicates that $\eta^\pm$ can be expressed as
\begin{align*}
\eta^\pm = \frac{2RT}{F}\sinh^{-1}\left(\frac{F}{2i_0^\pm}j_n^\pm\right).
\end{align*}
The terminal voltage $V$ is
\begin{align}\label{SPM-voltage}
\nonumber
    V_{\mathrm{SPMT}}(t) &= U^+(c_{ss}^+(t))-U^-(c_{ss}^-(t))+\eta^+ -\eta^- \\ & \quad  - \left(\frac{R_f^+}{a_s^+L^+}  + \frac{R_f^-}{a_s^-L^-}\right)I(t) ,
\end{align}
where $U^+$ and $U^-$ are the equilibrium potentials, and $R_f^+$ and $R_f^-$ are the solid-electrolyte interphase film resistances.

The charging/discharging of LiBs is accompanied by the heat generation and transfer. The change in temperature can be intense at large currents and notably affects the lithium-ion diffusion and electrochemical kinetics. Here, the temperature dependence of $D_s^\pm$ and $k^\pm$ is governed by the Arrhenius law:
\begin{align} \label{Arrhenius-Law}
    \psi = \psi_{\mathrm{ref}}\exp\left[\frac{E_{\psi}}{R}\left(\frac{1}{T_{\mathrm{ref}}}-\frac{1}{T(t)}\right)\right] ,
\end{align}
where $\psi$ is the $D_s^\pm$ or $k^\pm$, $T(t)$ is the lumped temperature, $R$ is the universal gas constant, and $E_{\psi}$ is the activation energy.
Based on the energy balance principle, the change of $T(t)$ is assumed to follow
\begin{align}
    \rho_{\mathrm{avg}} c_p \frac{dT(t)}{dt} = \dot{q}_{\mathrm{gen}} + \dot{q}_{\mathrm{conv}} ,
\end{align}
where $\rho_{\mathrm{avg}}$ is the cell bulk density, $c_p$ is the lumped specific heat capacity, $\dot{q}_{\mathrm{gen}}$ denotes the heat generation rate due to ohmic and entropic heating, and $\dot{q}_{\mathrm{conv}}$ is the convective heat removal rate with the ambience. Further, $\dot{q}_{\mathrm{gen}} $ and $\dot{q}_{\mathrm{conv}}$ are given by
\begin{align*}
\dot{q}_{\mathrm{gen}} &= I(t)\left[V(t)-(U^+(\Bar{c}_{s}^+(t)) - U^-(\Bar{c}_{s}^-(t)))\right] \\ &\quad + I(t)T(t)\frac{\partial}{\partial T}\left[U^+(\Bar{c}_{s}^+(t)) - U^-(\Bar{c}_{s}^-(t))\right] ,\\
\dot{q}_{\mathrm{conv}} &= -h_{\mathrm{cell}}\left(T(t)-T_{\mathrm{amb}}(t)\right) ,
\end{align*}
where $T_{\mathrm{amb}}$ is the ambient temperature, $h_{\mathrm{cell}}$ is the convective heat transfer coefficient, and the bulk concentration $\Bar{c}_{s}^\pm(t)$ is given by:
\begin{align*}
\Bar{c}_{s}^\pm(t) = \frac{3}{(R_s^\pm)^3} \int_{0}^{R_s^\pm} r^2 c_s^\pm(r,t) dr .
\end{align*}
We define the anodic surface SoC and bulk SoC as
\begin{align}\label{SPM-SoC}
    \mathrm{SoC}_{\mathrm{surf}} = \frac{c_{ss}^-(t)}{c_{s,\max}^-},
    \ \
    \mathrm{SoC}_{\mathrm{bulk}} = \frac{\Bar{c}_{s}^-(t)}{c_{s,\max}^-}.
\end{align}

\begin{figure}[t!]
\centering
\subfloat[][]{
    \includegraphics[width = .49\textwidth,trim={6.5cm 2cm 5.7cm 6.5cm},clip]{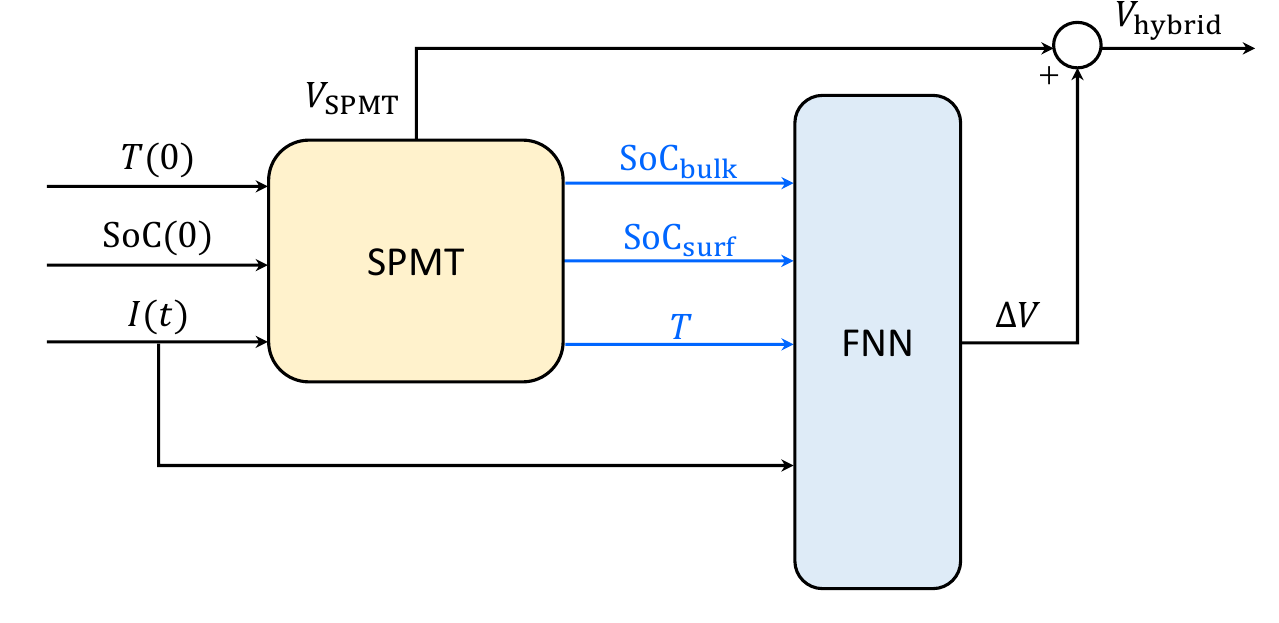}
    \label{Fig.: SPMTNet-1}}
    
\subfloat[][]{
    \includegraphics[width = .49\textwidth,trim={4.5cm 4cm 8.5cm 6cm},clip]{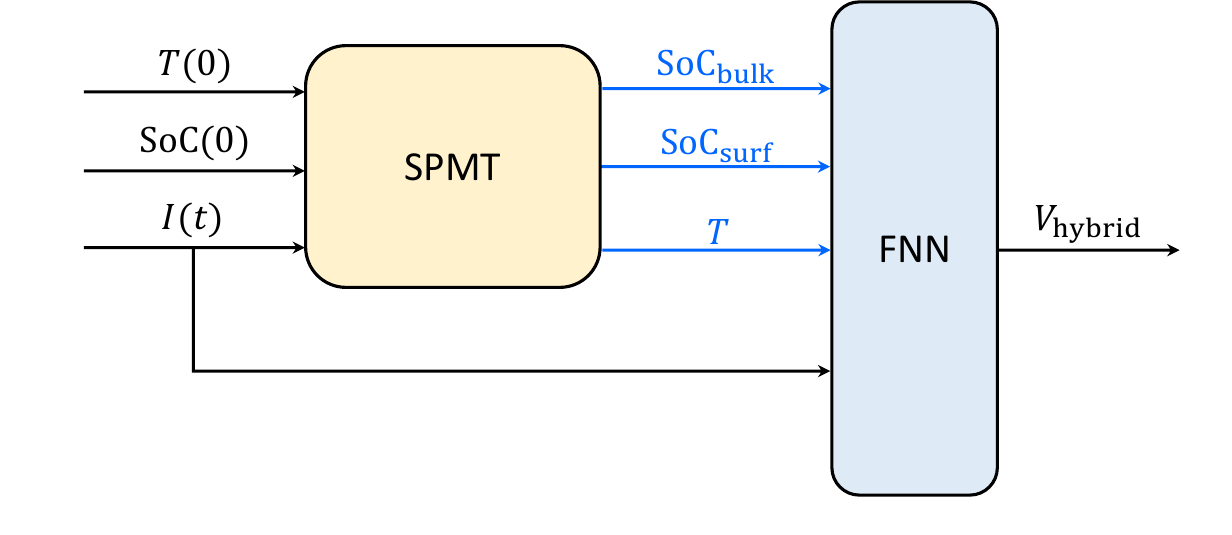}
    \label{Fig.: SPMTNet-2}}
\caption{Block diagrams of (a) the SPMTNet-1 model and (b) the SPMTNet-2 model.}
\end{figure}

\begin{table}[t!]
  \ra{1.2}
  \centering
  \renewcommand{\arraystretch}{}
  \subfloat[][]{
  \resizebox{\columnwidth}{!}{
  \begin{tabular}{c|c|c c c}
    \hline
    \toprule
     & \makecell{Current \\ profile} & \makecell{RMSE \\ (SPMT)} & \makecell{RMSE \\ (SPMTNet-1)} & \makecell{RER \\ (\%)} \\ \hline
     \multirow{9}{*}{\rotatebox[origin=c]{90}{Training}}
     
     & 0.2 C & 5.80 mV & 2.53 mV & 56.38 \\

     & 1 C & 20.34 mV & 4.29 mV & 78.91 \\

     & 2 C & 31.80 mV & 6.23 mV & 80.41 \\

     & 4 C & 62.48 mV & 5.90 mV & 90.56 \\

     & 6 C & 106.38 mV & 4.64 mV & 95.64 \\

     & 8 C & 157.58 mV & 4.24 mV & 97.31 \\

     & 10 C & 212.65 mV & 4.93 mV & 97.68 \\

     & US06 & 30.18 mV & 9.51 mV & 68.49 \\

     & LA92 & 23.54 mV & 9.83 mV & 58.24 \\\hline
     
     \multirow{7}{*}{\rotatebox[origin=c]{90}{Testing}}
     & 0.5 C & 11.12 mV & 4.65 mV & 58.18 \\

     & 3 C & 44.91 mV & 9.06 mV & 79.83 \\

     & 5 C & 83.31 mV & 4.99 mV & 94.01 \\

     & 7 C & 131.25 mV & 5.46 mV & 95.84 \\

     & 9 C & 184.78 mV & 4.61 mV & 97.51 \\

     & UDDS & 27.68 mV & 10.23 mV & 63.04 \\

     & SC04 & 26.27 mV & 8.82 mV & 66.43  \\ \hline
     
     \bottomrule
     \end{tabular}
     }
     \label{Table: SPMTNet-1}}
\quad

  \subfloat[][]{
  \resizebox{\columnwidth}{!}{
  \begin{tabular}{c|c|c c c}
    
    \hline
    \toprule
     
     & \makecell{Current \\ profile} & \makecell{RMSE \\ (SPMT)} & \makecell{RMSE \\ (SPMTNet-2)} & \makecell{RER \\ (\%)} \\ \hline
     
     \multirow{9}{*}{\rotatebox[origin=c]{90}{Training}}
     
     & 0.2 C & 5.80 mV & 2.86 mV & 50.69 \\

     & 1 C & 20.34 mV & 3.36 mV & 83.48 \\

     & 2 C & 31.80 mV & 5.56 mV & 82.52 \\

     & 4 C & 62.48 mV & 4.55 mV & 92.72 \\

     & 6 C & 106.38 mV & 3.73 mV & 96.49 \\

     & 8 C & 157.58 mV & 3.81 mV & 97.58 \\

     & 10 C & 212.65 mV & 3.41 mV & 98.40 \\

     & US06 & 30.18 mV & 10.71 mV & 64.51 \\

     & LA92 & 23.54 mV & 7.17 mV & 69.54 \\ \hline
     
     \multirow{7}{*}{\rotatebox[origin=c]{90}{Testing}}
     & 0.5 C & 11.12 mV & 5.07 mV & 54.41 \\

     & 3 C & 44.91 mV & 6.03 mV & 86.57 \\

     & 5 C & 83.31 mV & 4.38 mV & 94.74 \\

     & 7 C & 131.25 mV & 3.49 mV & 97.34 \\

     & 9 C & 184.78 mV & 4.38 mV & 97.63 \\

     & UDDS & 27.68 mV & 8.73 mV & 68.46 \\

     & SC04 & 26.27 mV & 9.77 mV & 62.81 \\ \hline
     
     \bottomrule
     \end{tabular}
     }
     \label{Table: SPMTNet-2}}
     \caption{Training/testing performance of (a) the SPMTNet-1 model and (b) the SPMTNet-2 model under different current profiles, in comparison with the SPMT model.}
\end{table}

\begin{figure*}[t!]
    \centering
    \subfloat[][Testing results of the SPMTNet-1 model under 0.5/3/7 C constant-current discharging. Marker symbols: line ``$-$'' for DFN;  circle ``$\circ$'' for SPMT; plus ``$+$'' for SPMTNet-1.]{
      \centering
      \includegraphics[width = .46\textwidth,trim={.8cm 0cm 1cm .7cm},clip]{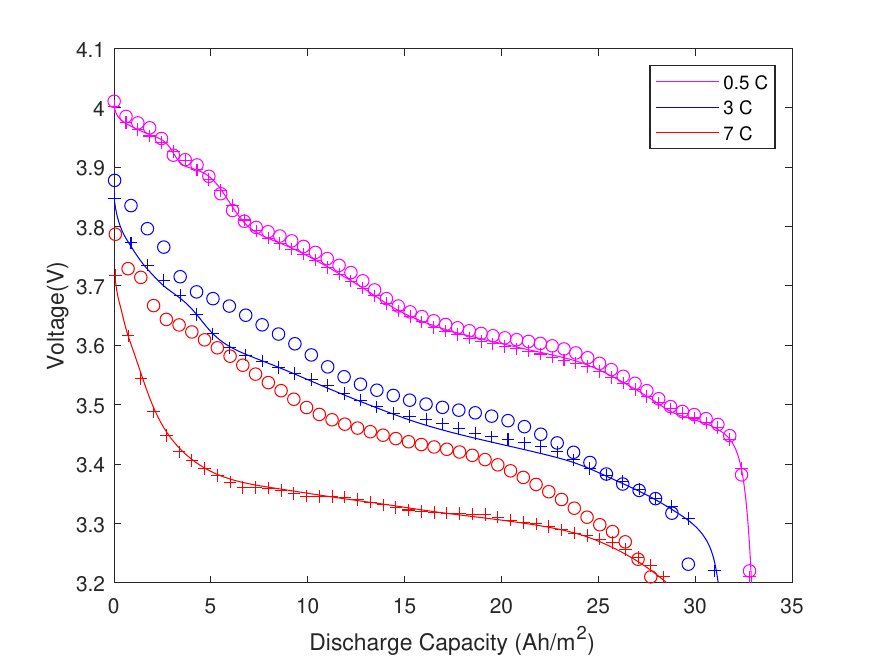}
      \label{HYBRID-I-test-constant-current}}

\subfloat[][Testing results of the SPMTNet-1 model under SC04 discharging.]{
    \centering
    \includegraphics[width = .46\textwidth,trim={.8cm 1.2cm 1cm 1.4cm},clip]{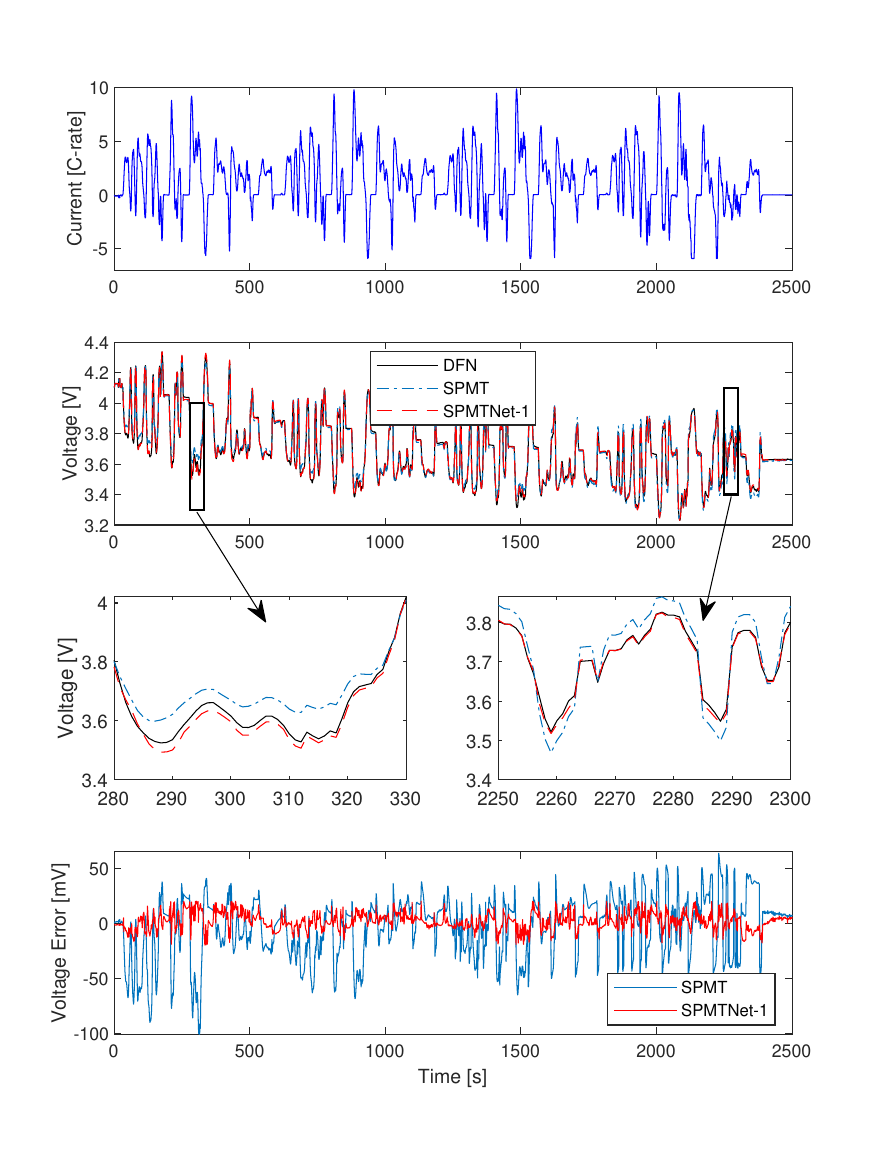}
    \label{HYBRID-I-test-UDDS-B}}
\quad
\subfloat[][Testing results of the SPMTNet-2 model under discharging by the UDDS profile.]{
    \centering
    \includegraphics[width = .46\textwidth,trim={.8cm 1.2cm 1cm 1.4cm},clip]{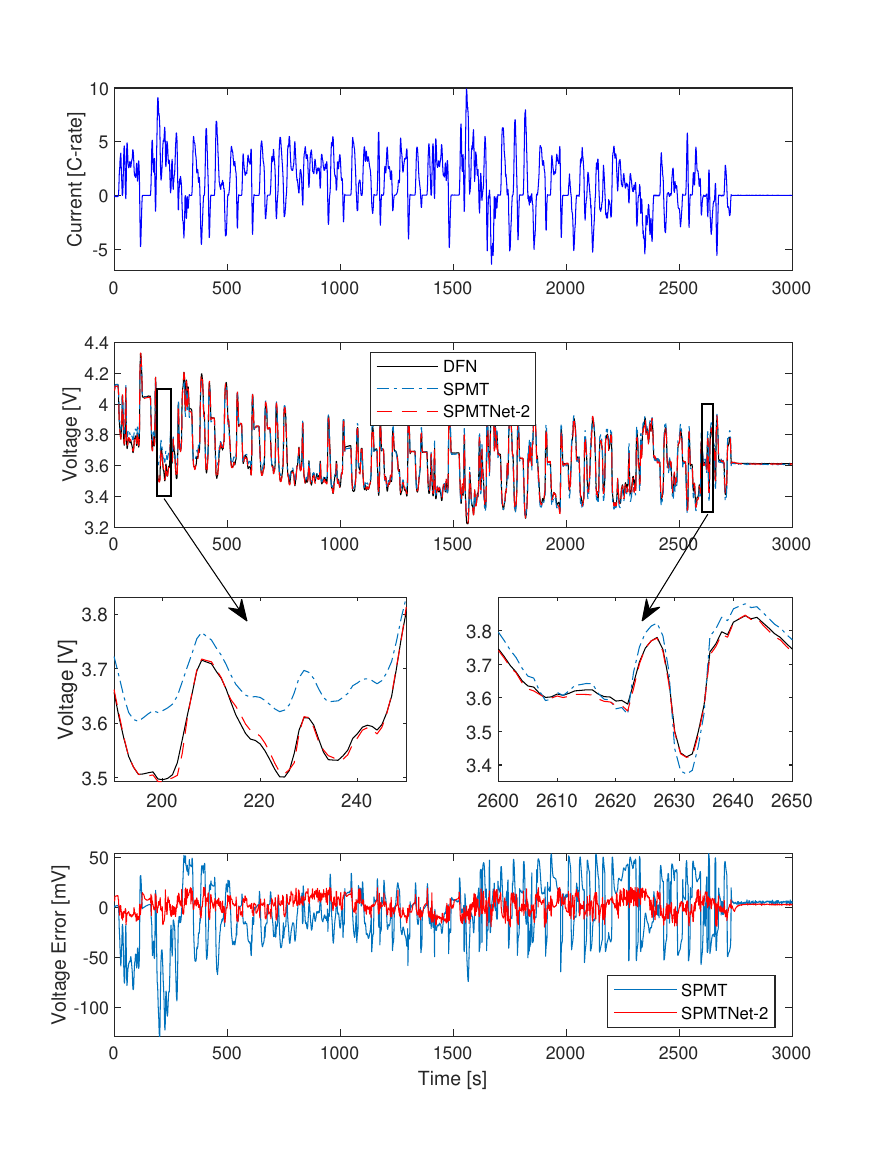}
    \label{HYBRID-II-test-US06}}
\caption{Testing results of the SPMTNet-1 and SPMTNet-2 models.}
\end{figure*}

Summarizing~\eqref{SPM-Diffusion}-\eqref{SPM-SoC}, we obtain a complete formulation of the SPMT model. This model is among the most computationally fast electrochemical models. It can offer good accuracy when low to medium currents are applied. However, its prediction performance at high C-rates or in the presence of uncertainty will deteriorate seriously, due to some  simplifications inherent to it.

Building upon the HYBRID-1 and HYBRID-2 frameworks, we propose SPMTNet-1 and SPMTNet-2, with their structures shown in Figs.~\ref{Fig.: SPMTNet-1} and~\ref{Fig.: SPMTNet-2}. These two hybrid models both combine the SPMT model with an FNN. SPMTNet-1 is designed to capture the residual between the actual voltage and the SPMT's prediction, and SPMTNet-2 is made to approximate the terminal voltage. For both, the FNN takes $\mathrm{SoC}_\mathrm{bulk}$, $\mathrm{SoC}_\mathrm{surf}$ and $T$ derived from the SPMT model as its input variables, leveraging an awareness of the physical model's state to make prediction.

\begin{remark}
The choice is non-unique for the variables used to represent the SPMT's electrochemical state and fed to the FNN. For instance, an expedient way is to just use the full electrochemical state of the SPMT. This, however, will cause extremely high training and computational costs. Our study shows that just several simple, aggregated state variables will suffice. This feature in effect reduces demands for training data and computation considerably, making the proposed hybrid modeling frameworks amenable to practical applications. After much trial-and-error search, we identify that the pair of
$\mathrm{SoC}_{\mathrm{bulk}}$, $\mathrm{SoC}_{\mathrm{surf}}$ and $T$ is a favorable choice for SPMTNet-1 and SPMTNet-2 in terms of both computational efficiency and prediction performance.
\end{remark}

\subsection{Simulation Validation}
\label{sec:SPMT hybrid simulation}

We performed simulation to validate the effectiveness of the proposed SPMTNet-1 and SPMTNet-2 models. The simulation settings are as follows:
\begin{itemize}

\item The DFN model with thermal dynamics, which is acknowledged as a generic and reliable electrochemical-thermal model, was used as the benchmark to assess the SPMTNet-1 and SPMTNet-2 models.  

\item We used parameters from the DUALFOIL simulation package~\cite{DualFoil} to run the DFN model representing an LCO/graphite battery that operates between 4.1 and 3.2 V to generate synthetic data as the ground truth.

\item The synthetic data were divided into the training and test datasets. The training datasets were produced by applying constant discharging currents at 0.2/1/2/4/6/8/10 C and variable currents created based on the US06 and LA92 driving cycles~\cite{EPA}. The test datasets were obtained by applying constant discharging currents at 0.5/3/5/7/9 C and variable currents created based on the UDDS and SC04~\cite{EPA}. 
Here, all variable current profiles were scaled to a maximum current of around 10 C. In all cases, the initial temperature $T(0)=T_\mathrm{amb}=25^\circ$C. 

\item  Both the SPMTNet-1 and SPMTNet-2 models employ a four-layer FNN as shown in Fig.~\ref{fig:NN}. Each of the two hidden layers has 32 neurons. The input and output of the FNN are as specified in Section~\ref{sec: The SPMTNets}. The rectified linear unit (ReLU) function was chosen as the activation function for the two hidden layers, and the linear activation function chosen for the output layer. Keras, a Python-based deep learning library, was used to create, train  and implement the FNN.  Because the magnitudes of the FNN's input variables vary across different orders of magnitude, the input data were pre-processed by normalization, as often recommended in the practice of NNs. 

\item We utilized the root-mean-square error as a metric to evaluate a model's performance:
\begin{align*}
\mathrm{RMSE} = \sqrt{\frac{1}{N }\sum_{i=1}^{N } \left(V_{\mathrm{true},i} -V_{\mathrm{model},i}  \right)^2},
\end{align*}
where $V_{\mathrm{true},i}$ is the true voltage at time $i$,  $V_{\mathrm{model},i}$ is the model-based voltage prediction, and $N$ is the total number of data points. Furthermore, a relative error reduction ($\mathrm{RER}$) was introduced to quantify the improvement of the SPMTNet-1 and SPMTNet-2 over the SPMT, which is defined as
\begin{align*}
   \mathrm{RER} = \frac{\mathrm{RMSE_{SPMT}}-\mathrm{RMSE_{SPMTNet}}}{\mathrm{RMSE_{SPMT}}} \times 100\%.
\end{align*}
\end{itemize}

We began with validating the SPMTNet-1 model. Table~\ref{Table: SPMTNet-1} summarizes its performance over all the training datasets and compares it with the baseline SPMT model. We observed that the SPMTNet-1 model offers remarkable accuracy in all training scenarios. It consistently outperforms the SPMT  model by a considerable margin, especially at medium to very high currents.  Further, we applied the trained SPMTNet-1 model to the test datasets to appraise its prediction performance. Table~\ref{Table: SPMTNet-1} shows a quantitative evaluation, and Figs.~\ref{HYBRID-I-test-constant-current}-\ref{HYBRID-I-test-UDDS-B} demonstrate a visual assessment in the cases of constant 0.5/3/7 C and the SC04 profiles. As is seen, the SPMTNet-1 still retains high accuracy in the testing phase, proving its strong predictive ability.

For the SPMTNet-2 model, Table~\ref{Table: SPMTNet-2} further shows its training/testing performance across different test datasets, and Fig.~\ref{HYBRID-II-test-US06} displays its prediction under the UDDS-based test dataset. These results show that the SPMTNet-2 is also greatly effective in grasping and forecasting the dynamics of LiBs.

Finally, we emphasize that the SPMTNet-1 and SPMTNet-2 models provide higher testing accuracy and better   voltage  predictive performance than the existing hybrid models for LiBs, e.g.,~\cite{Park:ACC:2017}, as extensive simulation reveals. This underscores the efficacy of the proposed design that feeding a physics-based model's state information into the ML model.

\section{Hybrid Modeling {\em via} NDC+FNN}
\label{sec:ECM Hybrid}
Section 3 shows the effectiveness of integrating electrochemical modeling with ML for modeling of LiBs. A subsequent question of interest is whether we can integrate ECMs with ML based on the proposed HYBRID-1 and HYBRID-2 frameworks. ECMs have simplistic structures and fast computation, and hybrid models based on them can be beneficial for various real-world battery management tasks. In this section, we blend the NDC model, an ECM developed recently in~\cite{Ning:TCST:2020,Tian:IECON:2018}, with an FNN to develop two hybrid models, named NDCNet-1 and NDCNet-2, respectively, and experimentally investigate their performance.

\subsection{The  NDCNet-1 and NDCNet-2 Models}


\begin{figure}[t!]
\centering
\subfloat[][]{
    \centering
    \includegraphics[width = .5\textwidth,trim={6.5cm 7cm 5.7cm 3.9cm},clip]{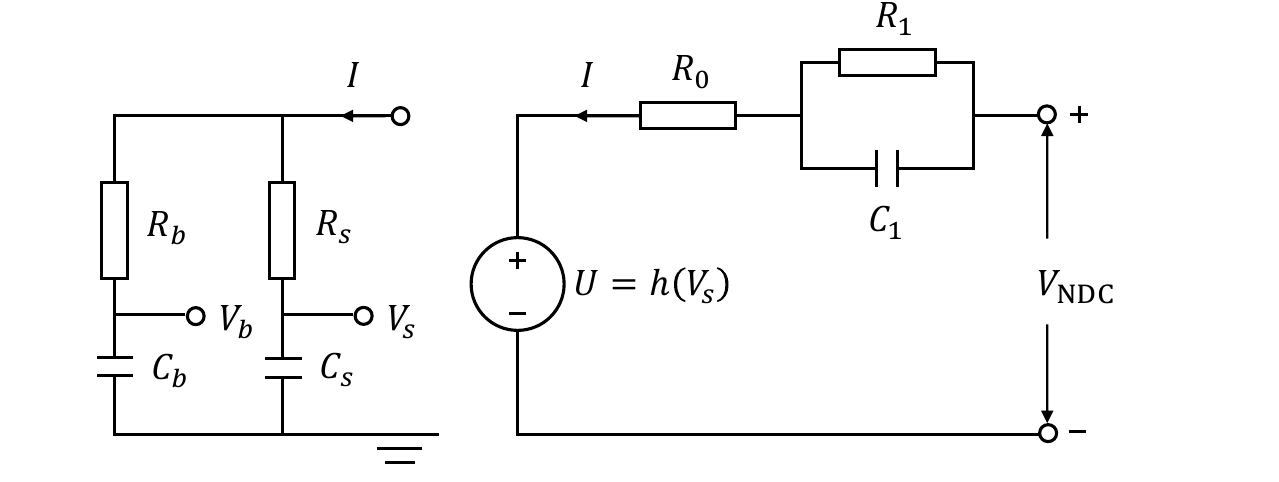}
    \label{fig: NDC}}

\subfloat[][]{
    \centering
    \includegraphics[width = .49\textwidth,trim={6.5cm 2.05cm 5.7cm 6.5cm},clip]{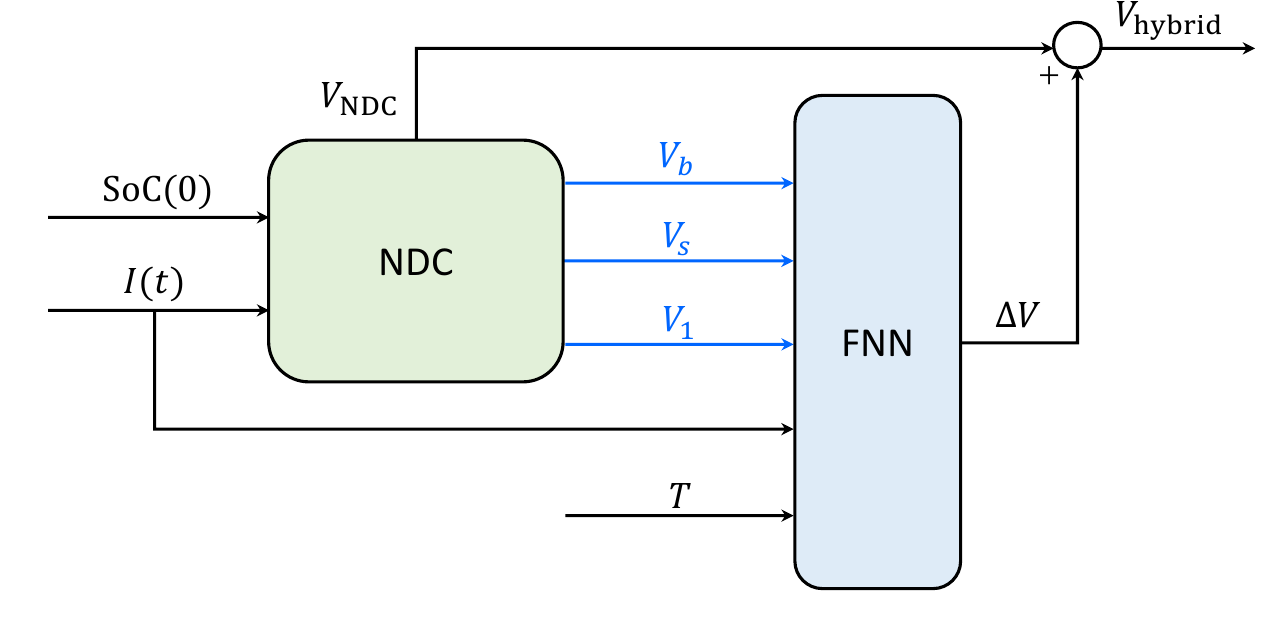}
    \label{fig: HYBRID3}}
    
\subfloat[][]{
    \centering
    \includegraphics[width = .49\textwidth,trim={4.5cm 4cm 8.5cm 6cm},clip]{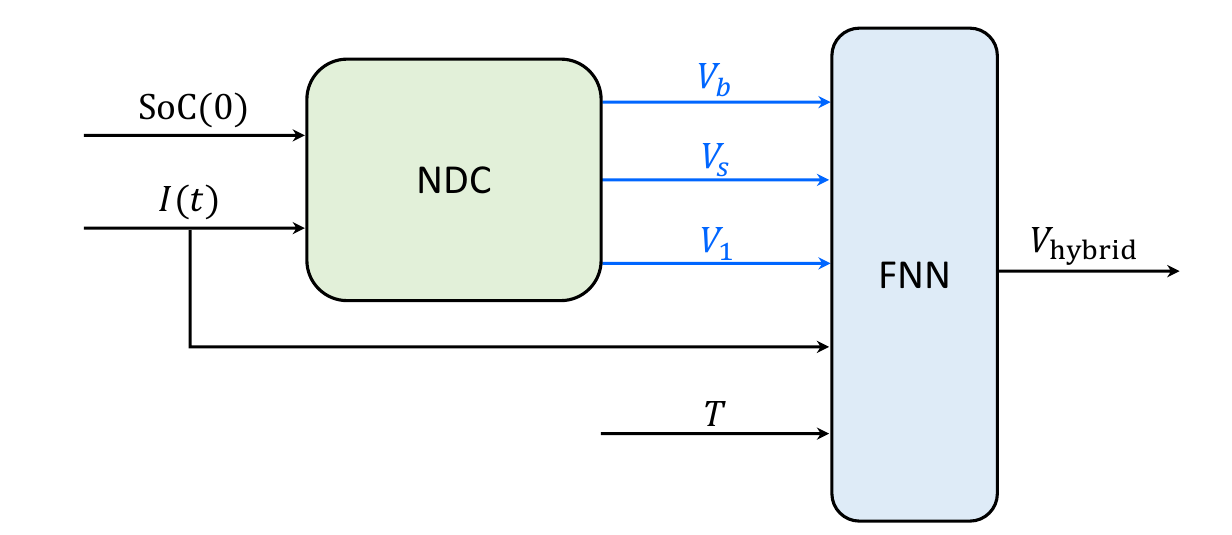}
    \label{fig: HYBRID4}}
\caption{(a) The NDC model, (b) the NDCNet-1 model and (c) the NDCNet-2 model.}
\end{figure}

The NDC model maps the diffusion and electrical processes in a LiB cell to a circuit of electrical components. As shown in Fig.~\ref{fig: NDC}, the circuit includes two coupled parts. The first (left) part simulates the diffusion in the cell's electrode, which comprises two R-C pairs, $R_b$-$C_b$ and $R_s$-$C_s$, configured in parallel. The $R_b$-$C_b$ analogously represents the bulk inner region of the electrode, and the $R_s$-$C_s$ represents the surface region of the electrode. As such, $C_b \gg C_s$ and $R_b \gg R_s$, where $R_s$ can often be set as 0~\cite{Ning:TCST:2020}. The charge transfer between $C_b$ and $C_s$ mimics the diffusion of lithium ions in the electrode~\cite{Fang:TCST:2016}. The second (right) part simulates the dynamic output voltage of the battery, which consists of a voltage source $U$, a resistor $R_0$, and an R-C pair $R_1$-$C_1$ pair connected in series. Here, $U=h(V_s)$ plays the role of the open-circuit voltage source. In addition, the $R_1$-$C_1$ approximates the voltage transients caused by charge transfer on the electrode/electrolyte interface, and $R_0$ accounts for the ohmic resistance and solid electrolyte interface resistance.

The state-space equations of the NDC model are given by
\begin{subequations} \label{NDC-SS}
\begin{align}
\begin{bmatrix}
\dot{V_b}(t) \\ \dot{V_s}(t) \\ \dot{V_1}(t)
\end{bmatrix}
&= A
\begin{bmatrix}
V_b(t) \\ V_s(t) \\ V_1(t)
\end{bmatrix} 
+ BI(t),\\ 
V_{\mathrm{NDC}}(t) &= h(V_s(t)) - V_1(t) + R_0 I(t),
\end{align}
\end{subequations}
where $V_b$, $V_s$ and $V_1$ are the voltage across $C_b$, $C_s$ and $C_1$, respectively. Here, 
\begin{align*}
    A =
    \begin{bmatrix}
    \frac{-1}{C_b(R_b+R_s)} & \frac{1}{C_b(R_b+R_s)} & 0 \\
    \frac{1}{C_s(R_b+R_s)} & \frac{-1}{C_s(R_b+R_s)} & 0 \\
    0 & 0 & \frac{-1}{R_1C_1}
    \end{bmatrix}, 
    B =
    \begin{bmatrix}
    \frac{R_s}{C_b(R_b+R_s)}\\
    \frac{R_b}{C_s(R_b+R_s)}\\
    \frac{-1}{C_1}
    \end{bmatrix}.
\end{align*}
In this study, we parameterize the $h(V_s)$   as
\begin{align*}
    h(V_s) = \frac{\alpha_1 V_s^2 + \alpha_2 V_s + \alpha_3}{V_s^3 + \alpha_4 V_s^2 + \alpha_5 V_s + \alpha_6},
\end{align*}
where $\alpha_i$ for $i=1, 2, ... , 6$ are the coefficients. Further, we have $V_b=V_s=0$ V when the cell is depleted ($\mathrm{SoC} = 0\%$), and 
 $V_b=V_s=1$ V when the cell is fully charged ($\mathrm{SoC}=100\%$). The total charge capacity of the cell thus is $C_b+C_s$. Then, the SoC is given by
\begin{align} \label{NDC-SoC}
    \mathrm{SoC} = \frac{C_bV_b+C_sV_s}{C_b+C_s} \times 100\%.
\end{align}
Finally, the internal resistance $R_0$ is assumed to be SoC-dependent:
\begin{align} \label{NDC-Ro}
    R_0 = \gamma_1 + \gamma_2e^{-\gamma_3\mathrm{SoC}} + \gamma_4e^{-\gamma_5(1-\mathrm{SoC})}.
\end{align}

\begin{table}[t!]
  \ra{1.2}
  \centering
  \renewcommand{\arraystretch}{}
  \subfloat[][]{
  \resizebox{\columnwidth}{!}{
  \begin{tabular}{c|c|c c c}
    
    \hline
    \toprule
     
     & \makecell{Current \\ profile} & \makecell{RMSE \\ (NDC)} & \makecell{RMSE \\ (NDCNet-1)} & \makecell{RER \\ (\%)} \\ \hline
     \multirow{7}{*}{\rotatebox[origin=c]{90}{Training}}
     
     & 1 C & 20.47 mV & 3.56 mV & 82.61\\

     & 2 C & 68.77 mV & 5.05  mV & 92.66\\

     & 5 C & 194.67 mV & 5.17 mV & 97.34\\

     & 7 C & 274.75 mV & 4.35 mV & 98.42\\

     & 8 C & 318.85 mV & 5.62 mV & 98.24\\

     & US06 & 33.70 mV & 8.67 mV & 74.27\\

     & SC04 & 38.19 mV & 6.68 mV & 82.50\\ \hline

     \multirow{5}{*}{\rotatebox[origin=c]{90}{Testing}}
     
     & 3 C & 112.67 mV & 11.25 mV & 90.02\\

     & 4 C & 150.63  mV & 10.87 mV & 92.78\\

     & 6 C & 236.51 mV & 7.83 mV & 96.69\\

     & UDDS & 32.92 mV & 10.96 mV & 66.71\\

     & LA92 & 28.36 mV & 9.30 mV & 67.21\\ \hline
     
     \bottomrule
     \end{tabular}}
      \label{Table: NDCNet-1}}
\quad

\subfloat[][]{
  \resizebox{\columnwidth}{!}{
  \begin{tabular}{c|c|c c c}
    
    \hline
    \toprule
     
     & \makecell{Current \\ profile} & \makecell{RMSE \\ (NDC)} & \makecell{RMSE \\ (NDCNet-2)} & \makecell{RER \\ (\%)} \\ \hline
     \multirow{7}{*}{\rotatebox[origin=c]{90}{Training}}
     
     & 1 C & 20.47 mV & 3.96 mV & 80.65\\

     & 2 C & 68.77 mV & 4.80 mV & 93.02\\

     & 5 C & 194.67 mV & 5.24 mV & 97.31\\

     & 7 C & 274.75 mV & 2.77 mV & 99.00\\

     & 8 C & 318.85 mV & 4.08 mV & 98.72\\

     & US06 & 33.70 mV & 9.24 mV & 72.58\\

     & SC04 & 38.19 mV & 5.99 mV & 84.32\\ \hline

     \multirow{5}{*}{\rotatebox[origin=c]{90}{Testing}}
     
     & 3 C & 112.67 mV & 14.05 mV & 87.53\\

     & 4 C & 150.63  mV & 10.72 mV & 92.88\\

     & 6 C & 236.51 mV & 9.14 mV & 96.14\\

     & UDDS & 32.92 mV & 10.85 mV & 67.04\\

     & LA92 & 28.36 mV & 8.60 mV & 69.68\\ \hline
     
     \bottomrule
     \end{tabular}}
     \label{Table: NDCNet-2}}
     \caption{Training/testing performance of (a) the NDCNet-1 model and (b) the NDCNet-2 model under different current profiles, in comparison with the NDC model.}
\end{table}

\begin{figure*}[t!]
\centering
\subfloat[][Testing results of the NDCNet-1 model under discharging by the LA92 profile with the cooling fan off.]{
    \centering
    \includegraphics[width = .47\textwidth,trim={.8cm 1.2cm 1cm 1.4cm},clip]{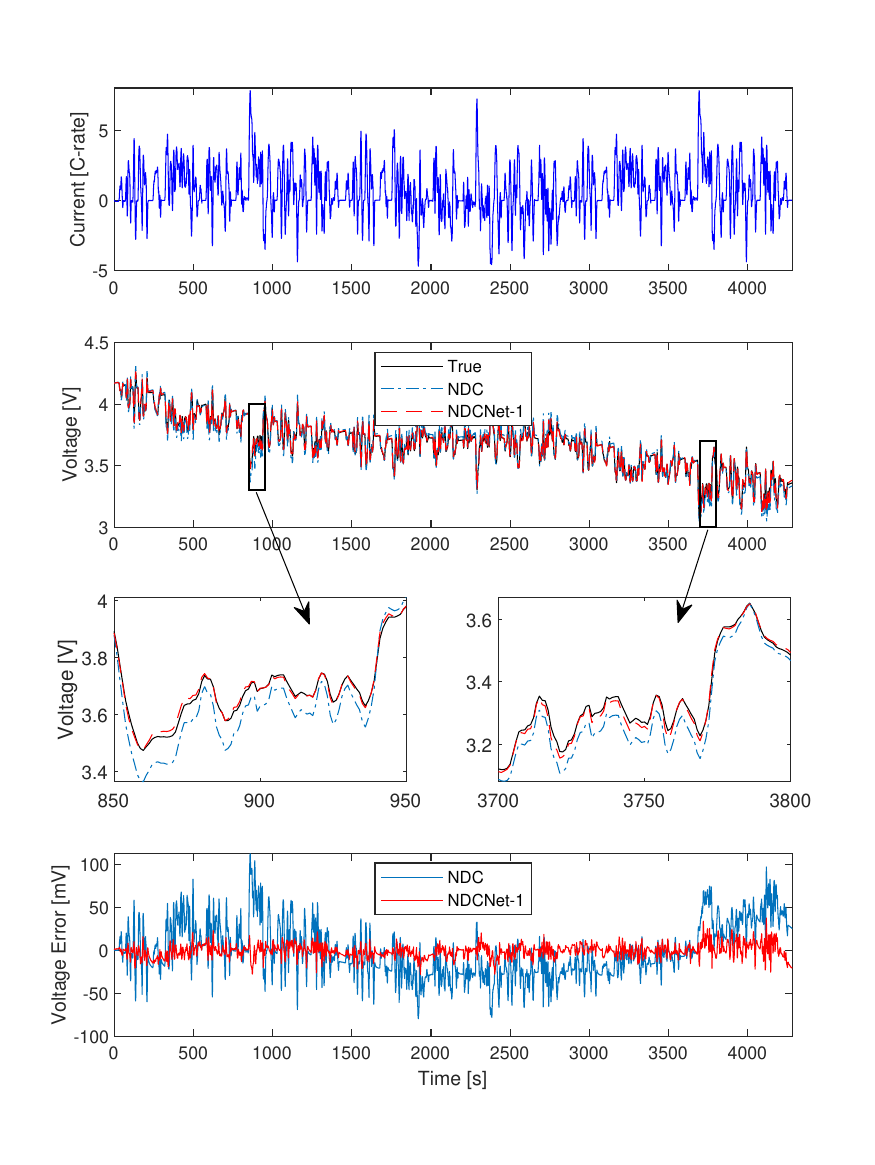}
    \label{HYBRID-III-test-LA92}}
\quad
\subfloat[][Testing results of the NDCNet-2 model under discharging by the UDDS profile with the cooling fan off.]{
    \centering
    \includegraphics[width = .47\textwidth,trim={.8cm 1.2cm 1cm 1.4cm},clip]{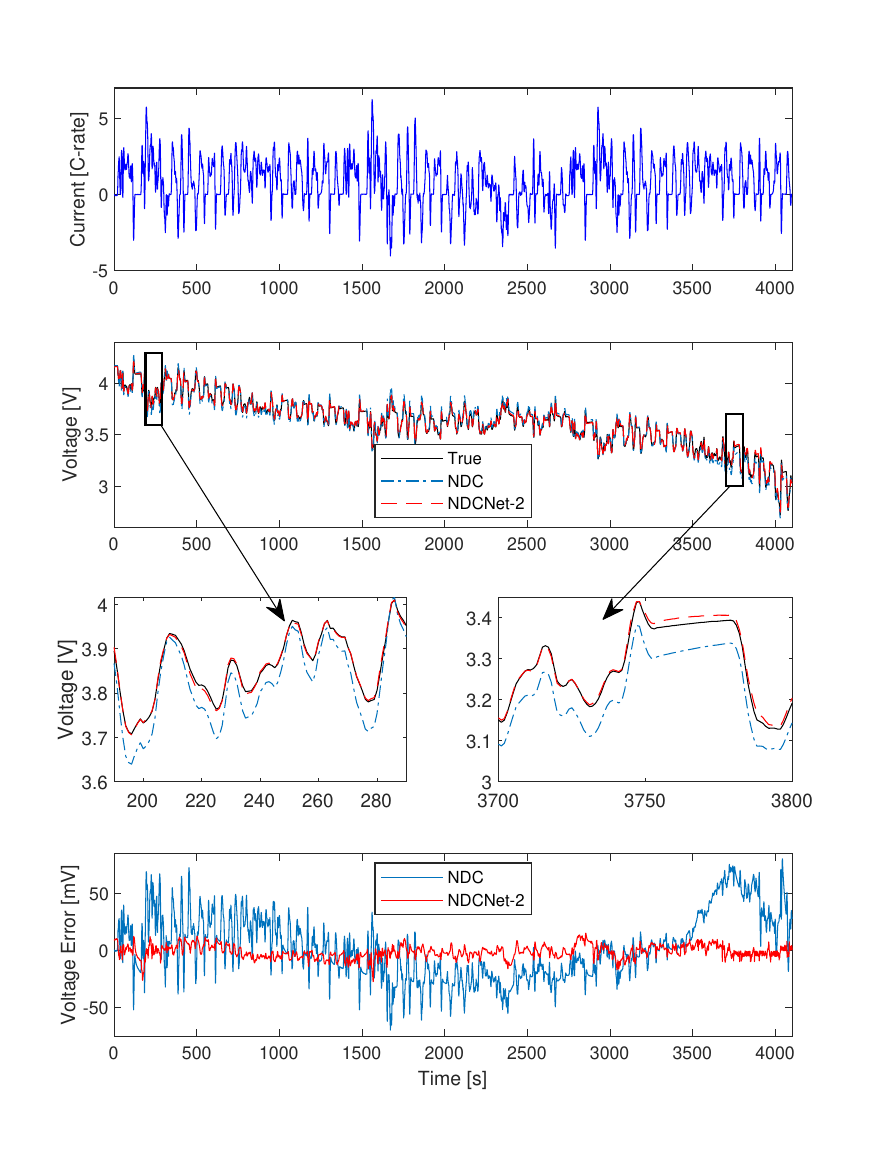}
    \label{HYBRID-IV-test-UDDS}}
\caption{Testing results of the NDCNet-1 and the NDCNet-2 models.}
\end{figure*}

The NDC model, {\color{blue}as summarized in~\eqref{NDC-SS}-\eqref{NDC-Ro},} simulates the charge diffusion in an electrode and the nonlinear voltage dynamics simultaneously. With this characteristic, it presents higher  voltage predictive accuracy at low to medium C-rates than earlier ECMs, including the Thevenin's model, and has found desirable use in SoC estimation and optimal charging~\cite{Movahedi:TTM:2021,Ning:TII:2021}. However, as with the SPM, its accuracy will decline at high C-rates. We are hence intrigued to develop NDC-based hybrid models to improve the  predictive performance.

We construct the NDCNet-1 and NDCNet-2, based on the HYBRID-1 and HYBRID-2 frameworks, respectively. Their structures are shown in Figs.~\ref{fig: HYBRID3}-\ref{fig: HYBRID4}. Here, the NDCNet-1 is designed to capture the NDC model's residual relative to the true terminal voltage directly, and the NDCNet-2 is made to learn and reproduce the terminal voltage. As the frameworks mandate, we feed the state variables of the NDC model, $V_b$, $V_s$ and $V_1$, to the FNN so that the FNN can perform physics-informed prediction. Besides, the  temperature $T$ is fed to the FNN so that the FNN can  capture the effect of the temperature in its voltage prediction.

\begin{figure}[t!]
\centering
\subfloat[][]{
    \centering
    \includegraphics[width = .47\textwidth,trim={.8cm .1cm 1.3cm .6cm},clip]{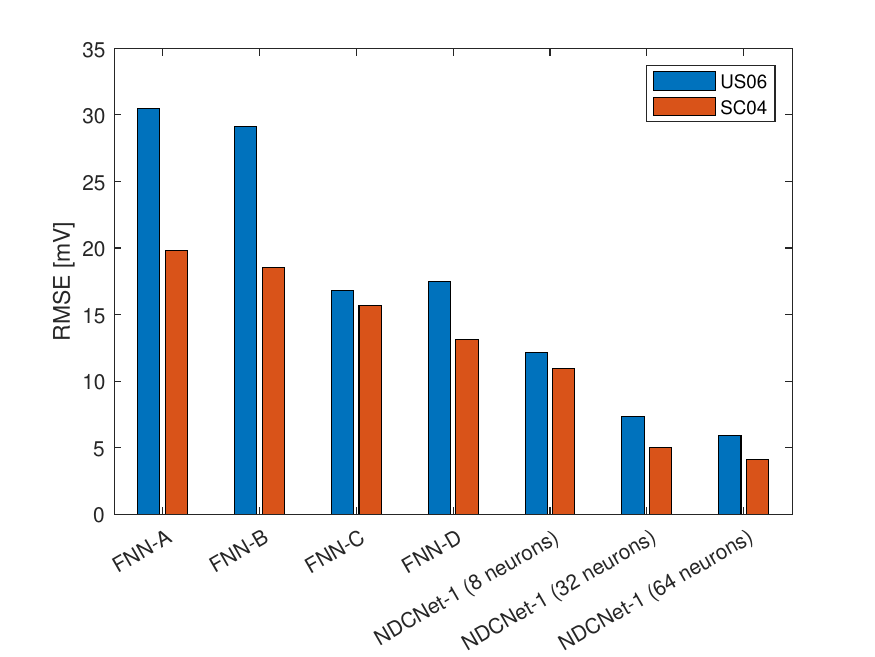}
    \label{PureMLTr}}
\quad

\subfloat[][]{
    \centering
    \includegraphics[width = .47\textwidth,trim={.8cm .1cm 1.3cm .6cm},clip]{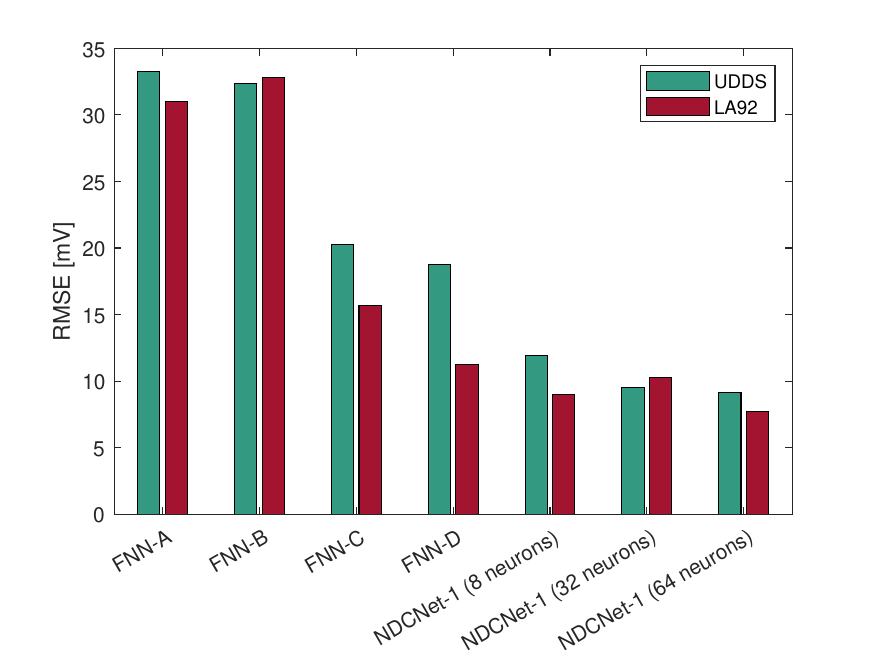}
    \label{PureMLTt}}
    \caption{(a) Training and (b) testing accuracy of pure FNN models and NDCNet-1 under variable current profiles when the cooling fan was off. The  numbers  of neurons in each hidden layer  are  shown in the parentheses.}
\end{figure}

\subsection{Experimental Validation}
\label{sec:NDC Hybrid Experiment}
We evaluated the proposed NDCNet-1 and NDCNet-2 models through experimental validation. The experimental settings are as follows:


\begin{itemize}

\item All the experimental data were collected on a Samsung INR18650-25R LiB cell using the PEC SBT4050 battery tester. The cell has a nominal capacity of 2.5 Ah and an operating range from 4.2 V to 2.8 V, with a maximum continuous discharging current of 20 A (8 C). 

\item The NDC model was extracted from experimental data using the parameter identification 1.0 approach in~\cite{Ning:TCST:2020}.

\item The training datasets were collected from experiments by applying constant discharging currents at 1/2/5/7/8 C and variable current profiles based on the US06 and SC04. The test datasets were based on constant discharging currents at 3/4/6 C and variable current profiles UDDS and LA92. Here, all variable current profiles were scaled to be between 0$\sim$8 C. The datasets were purposefully designed so as to validate the proposed models across low to very high C-rates.

\item In order to capture the influence of temperature, all types of current profiles were applied twice with an electric cooling fan on and off. The temperature was measured by a thermocouple attached to the cell's surface. During the experiments, the cell's temperature varied between 19$\sim$55$^{\circ}$C.

\item Both the NDCNet-1 and NDCNet-2 models adopt the same FNN architecture as in the SPMTNet-1 and SPMTNet-2. The performance metrics for evaluation are RMSE and RER as defined in Section~\ref{sec:SPMT hybrid simulation}.

\end{itemize}

The validation results of the NDCNet-1 and NDCNet-2 models are summarized in Tables~\ref{Table: NDCNet-1} and~\ref{Table: NDCNet-2}, respectively. Both models show considerable training accuracy---compared to the NDC model, they substantially decrease the prediction errors as measured by RMSE, especially when high C-rates are applied. The testing accuracy for both slightly declines but still remains high. Figs.~\ref{HYBRID-III-test-LA92}-\ref{HYBRID-IV-test-UDDS} further display the voltage prediction of the NDCNet-1 and NDCNet-2 models in comparison with the NDC model when the LA92 and UDDS profiles are applied. It is seen that the two  models consistently deliver much better prediction and, in particular, bring more performance enhancements at large currents.  These results demonstrate the NDCNet-1 and NDCNet-2 models as effective and powerful for voltage prediction. Note that both models are more parsimonious in structure than the SPMTNet-1 and SPMTNet-2 models, due to the simplicity of the NDC model. This makes them potentially more amenable to computation and real-world applications.

  Further, we  compared the NDCNet-1 model with pure FNN modeling to evaluate  our hybrid modeling design. Here, we   trained the NDCNet-1 model  with 8/32/64 neurons in each   hidden  layer   of  the FNN. The pure FNN models were designed to use the present and history information to predict the terminal voltage. They were set up as below:

\begin{itemize}

\item The FNN-A model: Input:  $I(k)$, $I(k-1)$ ,$T(k)$, $T(k-1)$, $\mathrm{SoC}(k)$, and $\mathrm{SoC}(k-1)$, where $k$ is the discrete time index, and SoC is based on Coulomb counting. Output:  $V(k)$. Structure:   two hidden layers with 128 neurons in each layer.

\item The FNN-B model: Input and output: the same as the FNN-A model.  Structure:   two hidden layers with 256 neurons in each layer.

\item The FNN-C model: Input:  $I(k)$, $I(k-1)$ ,$T(k)$, $T(k-1)$, $\mathrm{SoC}(k)$, $\mathrm{SoC}(k-1)$, and  $V(k-1)$. Output: $V(k)$. Structure:   two hidden layers with 128 neurons in each layer.

\item The FNN-D model:  Input and output: the same as the FNN-C model.  Structure: two hidden layers with 256 neurons in each layer.

\end{itemize}



Figs.~\ref{PureMLTr}-\ref{PureMLTt} illustrate the comparison results. We highlight two observations. First, all the versions of the NDCNet-1 model, despite having much smaller numbers of neurons and being trained on the same datasets, considerably outperform all the four pure FNN models in both training and testing. 
Second, pure ML models are prone to giving unreasonable predictions in testing scenarios. For instance, the FNN-D model has only slightly less accuracy than the  NDCNet-1 model when tested by the LA92 profile, but has much poorer performance under UDDS profile. 
The comparison shows that our hybrid modeling design can provide better prediction performance with simpler model structure  and offer good consistency in accuracy between training and testing.

\section{Aging-Aware Hybrid Modeling}
\label{sec:Aging Hybrid}

LiB cells age during cycling, which causes changes in material properties and affects the processes in charging/discharging~\cite{Krewer:JES:2018,Li:JPS:2021,Tang:AE:2019}. Aging manifests itself in capacity fade, internal resistance growth, and fast heat buildup. A LiB cell hence represents a time-varying system indeed. However, it has been found non-trivial to perform aging-aware LiB modeling, even though the problem has attracted some research. A main difficulty lies in characterizing physical relationships between aging and changes in a model's different parameters, which are often convoluted or elusive. Yet, the notion of hybrid modeling proposed in this paper can potentially allow to include an aging awareness into ML-based prediction, without tediously analyzing the underlying physics. To validate this promise, we further investigate aging-aware hybrid modeling in this section and focus on expanding the NDCNet-1 model,  with  similar results consistently obtained  for the other proposed hybrid models if they are modified in the same way. 

\begin{figure}[t!]
    \centering
    \includegraphics[width = 0.5\textwidth,trim={6.5cm 2.1cm 5.7cm 6.2cm},clip]{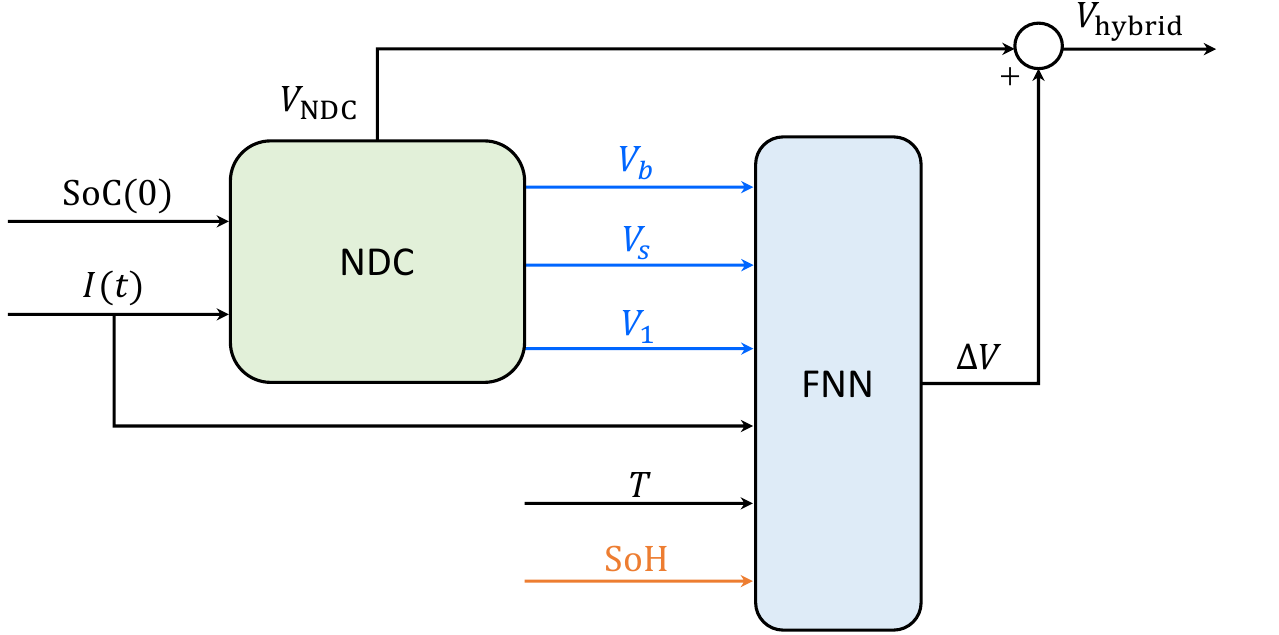}
    \caption{Block diagram of the AA-NDCNet-1.}
    \label{fig: HYBRID6}
\end{figure}

To quantify the aging condition, we consider SoH defined as the ratio between a cell's current capacity $Q_a$ and its initial capacity $Q_{\mathrm{init}}$:
\begin{align*}
    \mathrm{SoH} = \frac{Q_a}{Q_{\mathrm{init}}} \times 100\%.
\end{align*}
While SoH can be described in different ways, this definition suffices for our hybrid modeling, and its conciseness helps ensure  model parsimony.  Proceeding forward, we expand the NDCNet-1 model by including the above SoH. The new model is named as aging-aware NDCNet-1 (AA-NDCNet-1) and shown in Fig.~\ref{fig: HYBRID6}. It presents two main features. First, we apply SoH, which is calculated on a regular basis, as an additional input to the FNN. As such, the FNN becomes informed and aware of SoH when making prediction. Second, for the AA-NDCNet-1 model, we do not have to update the NDC model continually based on the aging condition. Instead, we solely use the FNN to capture the effect of SoH on the terminal voltage. This would bring significant convenience in practical use of the proposed model.

Next, we present the experimental validation of the AA-NDCNet-1 model. The experimental settings are as follows.

\begin{figure}[t!]
\centering
\subfloat[][]{
    \includegraphics[width = .5\textwidth,trim={1cm .7cm 1cm .8cm},clip]{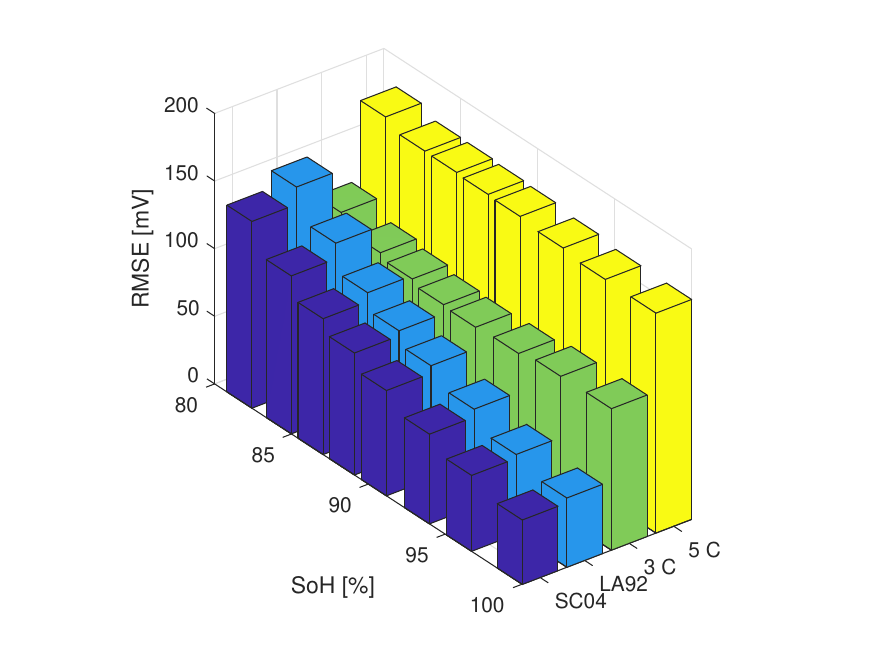}
    \label{AA-NDC}}
    
\subfloat[][]{
      \includegraphics[width = .5\textwidth,trim={1cm .7cm 1cm .8cm},clip]{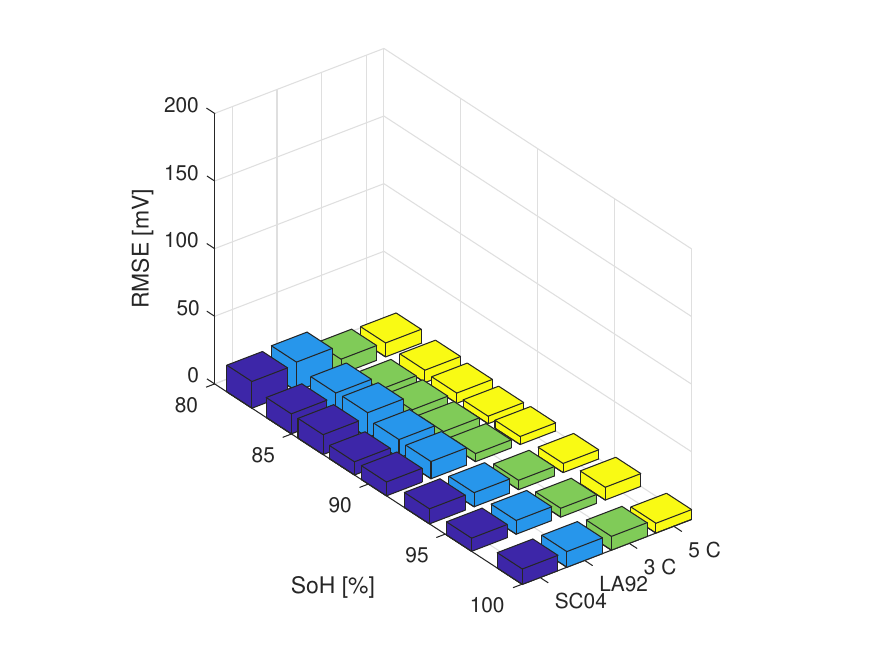}
      \label{AA-NDCNet}}
\caption{Comparison of (a) the NDC model and (b) the AA-NDCNet-1 model under different testing current profiles and SoH values.}
\end{figure}

\begin{itemize}

\item We collected the experimental data from two Samsung INR18650-25R LiB cells labeled as \#1 and \#2, respectively. Both cells underwent the same 450 charging/discharging cycles until their capacity $Q_a$ reached about 81\% of the initial capacity $Q_\mathrm{init}$. For a cycle, the cells were first charged based on the constant-current/constant-voltage charging protocol. Then, they were discharged by    repeatedly and periodically  applying the constant 1/2/3/4/5/6 C and UDDS/US06/LA92/SC04-based variable current profiles one after another. The cell's actual capacity $Q_a$ can be determined based on constant-current discharging at 4 C, as suggested by the cell's data sheet.

\item Cell \#1 was intended to generate training data, and cell \#2 was used to test the model. Here, we only used part of the datasets from cell \#1 to train the model; they included 1/2/4/6 C constant current profiles and variable current profiles based on the UDDS and US06 for SoH=81$\sim$100\%. The testing datasets from cell \#2 contained constant discharging current profiles at 3/5 C and variable current profiles based on the LA92 and SC04 for SoH = 81$\sim$100\%. All variable current profiles were scaled to be between 0 and 6 C in magnitude. The datasets span low to high C-rates in order to sufficiently assess the performance of the AA-NDCNet-1 model.

\item The NDC model was identified only once, using the data gathered from cell \#1 when SoH=100\%. The FNN hence aimed to capture the residual between $V_\mathrm{NDC}$ and $V_\mathrm{true}$ for different SoH values. For all scenarios, SoC  and C-rates  were calculated relative to the initial capacity $Q_\mathrm{init}$.

\item The AA-NDCNet-1 used the same FNN architecture as the NDCNet-1 in Section~\ref{sec:NDC Hybrid Experiment} and was evaluated by RMSE.

\end{itemize}

Figs.~\ref{AA-NDC}-\ref{AA-NDCNet} show the testing performance of AA-NDCNet-1 model compared with the NDC model when SoH = 99/95/93/90/88/86/84/81\%. Two observations are noteworthy. First, the RMSE of the identified NDC model steadily  increases  as the cell ages. However, the AA-NDCNet-1 model not only produces much smaller RMSE, almost consistently below 20 mV in all scenarios, but also preserves high accuracy throughout the aging process. Second, the AA-NDCNet-1 model, even though trained on cell \#1, is capable of well predicting the voltage behaviors of cell \#2 throughout its cycle life. This suggests the promise of making the AA-NDCNet-1 model a ``universal" hybrid model---one learned from a cell but widely applicable to other cells of the same type.

\section{Discussion} \label{Discussion}
Based on the results in Sections~\ref{sec:Electro Hybrid}-\ref{sec:Aging Hybrid}, we have the following remarks.
\begin{itemize}
    \item \textbf{Predictive accuracy}. The proposed catalog of hybrid models has shown not only high accuracy but also strong physical consistency in both simulations and experiments. We highlight that this merit rests on two factors. First, our hybrid models exploit physics-informed ML by feeding the state of a physical model into the FNN. The awareness of the physical model's status helps the FNN make better voltage  prediction. Second, data play a significant role in the overall prediction performance. Even though the proposed hybrid modeling frameworks reduce the dependence on data amounts compared to pure ML models, we still must use sufficient quantities of informative data to train the FNN. The data should effectively cover the spectrum of a LiB's dynamics and span the prediction ranges intended for the model in terms of C-rates, SoC, and SoH.
    
    \item \textbf{Computational efficiency}. For the proposed hybrid models, most of the computational burden comes from the FNN training. The training costs can vary, depending on the quantities of data and the structure of the FNN. However, we point out that the FNN employed in a proposed hybrid model can have a much simpler structure, compared to the case when a pure FNN is used for dynamic modeling of LiBs. This implies significantly lower computational costs in training. When deployed for online prediction, the hybrid models would allow fast computation. Our experiences showed the SPMTNet-1 and SPMTNet-2 models run much faster than the DFN model; further, the NDC-based hybrid models offer even higher computational efficiency.
    
    \item  \textbf{Prospective applications}.
As a main feature, the proposed hybrid frameworks and models are capable of making accurate voltage  prediction over wide  C-rate ranges. This makes them very useful for various kinds of practical LiB energy storage systems. In particular, they are suitable for LiB systems that must operate at high power conditions, for which accurate modeling  is still beyond the reach of today's  electrochemical models (e.g., the SPMT model) and ECMs (e.g., the Thevenin or NDC model) because of their either high computational costs or lack of accuracy. An application example in point is    electric  aircraft, which runs at up to 5 C in the take-off and landing phases~\cite{bills:arxiv:2021,YANG:Joule:2021}.  For such applications,  the proposed hybrid models can be used to  estimate the state-of-power and state-of-energy with their strong voltage prediction capabilities. This will be part of our subsequent work as an extension of  the presented study. The proposed models can also find promising use in LiB systems that operate at only low to medium C-rates. For such systems, a physical model with good fidelity is often easily available, but an FNN can be used to complement  the physical model to capture the uncertain nonlinear voltage. The resultant hybrid models will well lend themselves to voltage prediction and SoC estimation in this case.

    
    
    \item \textbf{FNN architecture selection}. A four-layer FNN, with the two hidden layers each having 32 neurons, was used throughout the model validation in the study, so as to assess the models on a common ground. However, we found out that other FNN architectures, e.g., one with fewer neurons in the hidden layers, can even lead to comparable performance. A practitioner may want to use an architecture to strike a balance between prediction accuracy and model size or complexity. This can be achieved by empirical tuning or deploying automated architecture search methods~\cite{Ren:ACM:2021}. Finally, our tries in the model validation with recurrent and long short-term memory NNs showed they can also be alternatives to FNNs here, though we choose the latter to present the study for their simplicity and tractability for practical application.
\end{itemize}

\section{Conclusion}
The ever-increasing adoption of LiBs across various sectors presents a pressing demand for accurate and computationally efficient models. In this paper, we proposed to integrate physics-based modeling with data-driven ML to meet this need. From this perspective, we developed the HYBRID-1 and HYBRID-2 modeling frameworks characterized by informing the ML model of the physical model's state information to significantly improve the voltage prediction performance and simplify the ML architecture.  We then applied the frameworks to investigate their viability in enabling effective integration of electrochemical models and ECMs with ML, respectively. We constructed four hybrid models, based on the notions of {\em SPMT+FNN} and {\em NDC+FNN}. We conducted extensive simulations and experiments to illustrate that all the four hybrid models can offer exceptionally high   voltage  predictive accuracy for LiBs operating at a wide range of C-rates. Further, we expanded the hybrid modeling design to embed an awareness of a LiB's aging condition into prediction, through making the ML informed of SoH. An NDC+FNN hybrid model was upgraded to achieve this end and experimentally validated to be capable of making accurate prediction under different SoH conditions. Our future work will include the application of the proposed models to different battery management tasks.

\section*{Acknowledgement}
This work was supported in part by National Science Foundation under Awards CMMI-1763093 and CMMI-1847651.

\balance
\bibliographystyle{elsarticle-num}
\bibliography{cas-refs}

\end{document}